%% file: History_rkellis.tex
\documentclass[a4paper,11pt]{article}
\pdfoutput=1 \usepackage{jheppub} \usepackage[T1]{fontenc} \usepackage{bm}
\usepackage{braket}
\usepackage[colorinlistoftodos]{todonotes}
\usepackage[utf8]{inputenc}
\DeclareUnicodeCharacter{0394}{\ensuremath{\Delta}}
\title{Perturbative QCD as a quantitative tool in the years 1976-2000\footnote{Presentation
  at the 4th International Symposium on the History of Particle Physics, 10--13/11/2025}}
\author{R. Keith Ellis,}
\emailAdd{keith.ellis@durham.ac.uk}
\affiliation{Institute for Particle Physics Phenomenology, Durham University, Durham, DH1 3LE, UK}
\preprint{IPPP/26/49}
\abstract{This paper traces the development of precision QCD in the years 1976-2000.
  This is after the
  discovery of asymptotic freedom, and after the exploration of the simplest
  processes based on the operator product expansion. The new theoretical tools
  of factorization, infra-red safety and resummation,
  needed to make predictions for the colliding beam machines of this era,
  are described. The role of computer algebra and modern spinor techniques
  for the calculation of amplitudes and cross sections are briefly reviewed.
  A selection of important processes calculated at
  next-to-leading order (or in limited cases beyond next-to-leading order) is presented.}
\begin{document}
\maketitle
\input body_rkellis.tex

\bibliography{History_rkellis.bib}
\bibliographystyle{unsrturl}
\end{document}

%% file: body_rkellis.tex
\def\as{\alpha_s}
\def\qb{\bar{q}}
\def\si{\sigma}
\def\ep{\epsilon}
\def\Tr{{\rm Tr}}
\def\tree{{\rm tree}}
\def\beq{\begin{equation}}
\def\eeq{\end{equation}}
\def\beqn{\begin{eqnarray}}
\def\eeqn{\end{eqnarray}}
\def\slsh{\rlap{$\;\!\!\not$}}
\def\db{{\dot{\beta}}}
\def\da{{\dot{\alpha}}}
\section{Setting the stage}
\subsection{QCD prior to 1976}
The point of departure for this talk is a world in which the property of asymptotic
freedom has been discovered,
\beq
\frac{d \as(Q^2)}{d \ln Q^2}= -b \alpha^2_s(Q^2)+O\big(\as^3(Q^2)\big)\,,\;\;\; b= \frac{33-2 n_f}{12 \pi}\, ,
\eeq
and Quantum Chromodynamics (QCD) has been proposed as a candidate
theory of the strong interactions. QCD has been used to justify
the approximate scaling observed in Deep-Inelastic Scattering (DIS), (albeit with logarithmic corrections).
The anomalous dimensions $\gamma(n)$,
which govern the scaling violations of the moments of the DIS structure functions, $M(n,Q^2)$,
have been calculated~\cite{Gross:1973zrg,Georgi:1974wnj} in leading order,
\beq \label{eq:moments}
\frac{d}{d \ln Q^2} M(n,Q^2) = \frac{\as (Q^2)}{2\pi} \gamma(n) M(n,Q^2) \, .
\eeq
This leads to logarithmic variation of the moments with $Q^2$, e.g.~for non-singlet moments,
\beq
M(n,Q^2)/M(n,\mu^2)= \Big[\frac{\as(\mu^2)}{1+\as(q^2)}\Big]^{\gamma(n)/(2 \pi b)}\, .
\eeq
The influence of strong interactions on nonleptonic weak interactions, a potential explanation for
the $\Delta I = \frac{1}{2}$ rule~\cite{Wilson:1970ag}, have also been calculated~\cite{Gaillard:1974nj,Altarelli:1974exa}.
It is important to note that all of these applications rely on the operator
product expansion. By late 1974 it was also accepted
that the $J/\psi$ particle was a triplet, spin one, charm-anticharm bound state,
whose narrow width could be explained by its mandatory decay into three gluons with a small QCD coupling as predicted by
asymptotic freedom~\cite{Appelquist:1974zd}.

The theoretical basis for this point of departure 
has been authoritatively reviewed in the previous incarnation of this conference~\cite{Hoddeson:1997hk},
especially by Bjorken in his talk {\it ``Deep-Inelastic Scattering: From Current Algebra to Partons''}
covering the years from 1969-1971 and by Gross in his review {\it ``Asymptotic Freedom and the Emergence of QCD''}.
Interventions by `tHooft and Veltman on {\it ``The path to renormalizability''} and Gell Mann on 
{\it ``Quarks, Color, and QCD''} also describe the advances made in the late 1960s and early 1970s.
In view of the focus of these previous historical reviews it is appropriate to begin this review in about 1976.
The middle of this decade was also my own personal point of departure, since my thesis~\cite{mythesis,Altarelli:1974ni}
submitted to the University of Oxford early in 1975 contained and exploited all the elements
of the standard model ($SU(3) \otimes SU_L(2)\otimes U(1)$), following the suggestions of my thesis advisors,
Guido Altarelli and Luciano Maiani.

\subsection{Slow acceptance of QCD}
The first thing to say, as emphasized also by Gross~\cite{Hoddeson:1997hk}, was that the acceptance of QCD
was not immediate throughout the physics community. In part this was due to the trepidation caused by
the assertion that QCD led to infra-red slavery or confinement, without any convincing proof.
In addition, because asymptotic
freedom was approached logarithmically, it was hard to find convincing experimental evidence
given the limited range of scales that was available. Also the energies being probed were
not asymptotic, so that the influence of hadronic scales could not be ignored.
Indeed as late as 1977, Marciano and Pagels~\cite{Marciano:1977su} wrote the tentative statement,
{\it ``In the last four years there has been a growing conviction among many theoretical physicists
that they have found a theory of the strong interaction..... Should this hope become
realized it would be a major triumph for theoretical physics.''} 

As other evidence of the relatively slow acceptance of QCD, I note
that in 1976 Feynman, Field and Fox~\cite{Field:1976ve} were attempting to
fit phenomenological forms for quark-quark scattering invariant
cross-sections. After reliable matrix elements for quark and gluon
scattering cross sections became available in September
1977~\cite{Combridge:1977dm}, they issued an
update~\cite{Feynman:1978dt} stating that,
{\it ``the quantum-chromodynamics approach is in reasonable accord with the
data although theoretical uncertainties (especially at low $p_T$)
make incontrovertible conclusions impossible at present. Crucial
tests of the theory require higher $p_T$ than are now available.''}

The task at hand in the period 1976-2000 was to confirm QCD as an acceptable theory of the strong interactions, by measurement of the
quantum numbers of the constituent quarks and gluons, and of the strong coupling. A more general aim was to search
for other constituents of the Standard model, such as the bottom quark (discovered in 1977), the top quark (discovered in 1995),
and the Higgs boson.
A number of new colliding beam machines were running during this period,
at CERN (ISR: 1971-1984, S$p\bar{p}$S: 1981-1991, LEP: 1989-2000),
at Fermilab (Tevatron: 1983-2011), at SLAC (SPEAR: 1972-1990, PEP: 1980-1986, SLC: 1989-1998) and at DESY (PETRA: 1978-1986, HERA: 1992-2007).
These machines required the development of new theoretical ideas to enable the use of perturbative
calculations in QCD. Although many calculation techniques followed the example of QED, the new ideas were unique to QCD.
These new ideas will be described in the next three sections.

\section{Factorization}
\label{factorization}
\subsection{Precursors to DGLAP}
Before the advent of asymptotic freedom a number of calculations showed that Bjorken scaling does
not hold without a transverse momentum cutoff.
Early calculations (without a transverse momentum cutoff),
such as by Chang and Fishbane~\cite{Chang:1970jk} with pseudo-scalar exchange
showed the dominance of ladder diagrams. Fishbane and Sullivan also investigated ladder diagrams in massive electrodynamics~\cite{Fishbane:1971mnh}.
Gribov and Lipatov~\cite{Gribov:1971zn,Gribov:1972ri,Gribov:1972rt} performed ladder calculations for both vector and pseudo-scalar meson theories.
Christ, Hasslacher and Mueller~\cite{Christ:1972ms} performed calculations in moment-space using the language 
of the light-cone expansion. These results were shown to be equivalent to the results for the scaling violations of Chang and Fishbane~\cite{Chang:1970jk}
and Gribov and Lipatov~\cite{Gribov:1971zn,Gribov:1972ri,Gribov:1972rt}.

The definitive results for QCD
in moment space were given by Gross and Wilczek~\cite{Gross:1973zrg}
and Georgi and Politzer~\cite{Georgi:1974wnj}, see Eq.~(\ref{eq:moments}).
Gross and Wilczek worked in part in the axial gauge as a method of banishing ghost operators.
\subsection{DGLAP}
One of the crucial steps in the development of the QCD improved parton model, was the
formulation of the scale dependence of the parton distributions directly in $x$ space.
This development is now associated with the names of Dokshitzer, Gribov, Lipatov, Altarelli and
Parisi (DGLAP)~\cite{Gribov:1971zn,Gribov:1972ri,Gribov:1972rt,Dokshitzer:1977sg,Parisi:1976qj,Altarelli:1977zs}.
The definitive papers for the case of QCD are the April 1977 results of
refs.~\cite{Altarelli:1977zs,Dokshitzer:1977sg}.
The result of ref.~\cite{Dokshitzer:1977sg} was an extension to QCD of earlier works on ladder
diagrams by Gribov and Lipatov~\cite{Gribov:1972ri,Gribov:1972rt}.

As we shall see the discovery of the DGLAP
equation, especially the work of Altarelli and Parisi, has important implications
for factorization.
The form of the DGLAP equation is,
\beq
\frac{d}{d \ln Q^2} f_i(x,Q^2) = \frac{\as (Q^2)}{2\pi} \int_x^1
\frac{d\xi}{\xi}\; P_{ij}(\xi) f_j\left(\frac{x}{\xi},Q^2\right) \; ,
\eeq
where $f_i(x,Q^2)$ is a scale-dependent parton distribution for a parton of type $i$.
Thus for example the splitting function of a quark into a quark is,
\beq
P_{qq}(x)=C_F \Bigg[ \frac{2}{[1-x]_{+}} -1-x +\frac{3}{2} \delta(1-x) \Bigg],\;\;\; C_F=\frac{4}{3}.
\eeq
The moments of the splitting functions $P_{ij}(x)$ are the anomalous dimensions~$\gamma_{ij}(n)$,
\beq
\gamma_{ij}(j) = \int_0^1 dx\; x^{n-1} P_{ij}(x) \; .
\eeq
In contrast to the moment distribution, the
$x$ space formulation has the advantage that the evolution requires knowledge of the parton distributions
only at values of $x/\xi$ which are greater than the target value of $x$.

Altarelli and Parisi do not quote the 
work of the Russian school, basing their treatment on iterative application of the Weisz\"acker-Williams
approximation, as suggested by an earlier work of Cabibbo and Rocca~\cite{Cabibbo:1974qb}, see also \cite{Baier:1973ms}.
The direct calculation of the splitting functions relies on the use of a physical gauge.
The treatment of ref.~\cite{Altarelli:1977zs} has the merit that the calculation of the splitting
probabilities does not require the specification of the physical process in which the partons
participate, which emphasizes their universal nature.
According to Parisi~\cite{Parisi:2025nob},{\it ``The crucial point was to shift the
  focus from Wilson operator expansion to the effective number of partons that was dependent on
  the resolution, i.e. $Q^2$. It was more than a computation: it was a change in the language we use.
The appropriate choice of language is one of the most important scientific tools."}

It is worth mentioning that the earlier 1976 paper of Parisi~\cite{Parisi:1976qj} contains the
lowest order QCD splitting probabilities, (although the probabilities given in that paper are not
$100\%$ accurate).
This paper also introduces the definition of the {\it plus distribution} to take into account virtual effects, 
\begin{equation} \label{plusdistribution}
\int_0^1 \, dx \, \frac{f(x)}{[1-x]_{+}}= \int_0^1 \, dx \, \frac{f(x)-f(1)}{(1-x)} \, .
\end{equation}
The plus distribution, similar to the Dirac delta function, is only defined under the integral sign. It is identical
to the normal function $1/(1-x)$ except at the point $x=1$.
The DGLAP splitting probabilities are inverse Mellin transforms of earlier results obtained by
Gross and Wilczek~\cite{Gross:1973zrg} and Georgi and Politzer~\cite{Georgi:1974wnj}.
Altarelli was concerned that the results in \cite{Altarelli:1977zs} might not be sufficiently novel to merit publication
and therefore took pains to include the splitting functions also for polarized targets.

\subsection{The parton model confronts QCD : Factorization \& Drell-Yan}
In this section I will describe the changes in the parton model required by QCD, in effect
the transformation of the na\"ive parton model~\cite{Feynman1972}
to the QCD parton model\footnote{The name na\"ive parton model
evidently irked Feynman -- on several occasions I heard him say, {\it ``Was it so na\"ive?''}}.
The first application of the QCD parton model beyond DIS was to lepton pair production.

Dilepton production in hadronic collisions
\beq
    {\rm hadron}~+~{\rm hadron} \to \gamma^* \to l^+ l^-~\big({\rm and~later}~ \to (Z \to \to l^+ l^-) ~{\rm or}~\to~(W \to l \nu)\big)
\eeq
was first investigated by Lederman et al.~\cite{Christenson:1970um,Lederman:1971pt} for muon pair production,
but named after Drell and Yan~\cite{Drell:1970wh} who wrote theoretical papers on the process in the parton model.  
Following the work of Altarelli and Parisi it was natural to ask whether the Drell-Yan
formula would hold, both at lowest order and beyond.
This topic was addressed by Politzer~\cite{Politzer:1977fi} and later also by Sachrajda~\cite{Sachrajda:1977mb}.
The key issue is whether all infrared sensitivity can be absorbed into the parton distribution functions,
both in electroproduction and in the Drell-Yan process. The first step was to show that it held to
order $g^2$. In order the regulate the infrared and collinear singularities
Politzer took the external quark and gluon legs slightly off-shell
and was able to show that the IR and collinear divergent terms were indeed
common between the Drell-Yan process and electroproduction.
Politzer also raised the question of whether factorization should hold to all orders.
\subsection{Dimensional regularization for collinear and IR singularities}
The extension of the off-shell regularization method to isolate the finite terms lying beneath the logarithms turned out to be a tricky procedure~\cite{Humpert:1979hb} and a cleaner method of calculation was needed.
The method of dimensional regularization, (i.e.~performing calculations in $d=4-2 \epsilon$ dimensions)
for the control of ultra-violet singularities, with the attendant benefits
for gauge theories of the preservation of gauge invariance became available
in 1972~\cite{tHooft:1972tcz,Bollini:1972ui}. Shortly thereafter the extension to
deal with IR singularities was proposed~\cite{Gastmans:1973uv}. Finally in 
1975 the extension to also regulate mass singularities~\cite{Marciano:1975de} was proposed.
The first calculation of NLO corrections in the factorization framework using dimensional
regularization was obtained in 1979~\cite{Altarelli:1979ub}. Other more-or-less successful
attempts to calculate the higher order corrections were made using a mass for the gluon~\cite{Kubar-Andre:1978eri}
as a regularization procedure and (following Politzer) by continuing
the external lines off-shell~\cite{Altarelli:1978id,Harada:1979bj,Abad:1978ke}.
The dimensional regularization procedure was so obviously superior
that all other regularization techniques were abandoned, and nowadays
no-one would dream of using any other regularization method.

\subsection{Phenomenological results for Drell-Yan}
The experimental data for the Drell-Yan process was found to lie above the prediction of the lowest order QCD parton model
by a factor of 2.2--2.4~\cite{Saclay-CERN-CollegedeFrance-EcolePoly-Orsay:1979yqw},
a discrepancy that was dubbed the K-factor by Altarelli~\cite{Altarelli:1979zh}.
The NLO corrections to the Drell-Yan formula~\cite{Altarelli:1979ub} were found to be large
at low $Q^2$ and approximately reproduced this K-factor enhancement, an important confirmation of the validity of QCD.
Taking a notional value of $\as/(2 \pi) \sim 1/20$ appropriate for $Q^2\sim 10$~GeV$^2$
the NLO correction is $150\%$ of the LO result. The size of the $\as$ correction would be higher (lower) for lower (higher) values of $Q^2$.
The QCD explanation of these results for Drell-Yan processes was sometimes questioned.
In the absence of the colour degree of freedom, the cross section would be a factor
of 3 larger, because with the coloured partons of QCD e.g.~a red quark can only annihilate with an anti-red quark.
Including the colour degree of freedom in QCD, it is found that the factor of 3 suppression is approximately
cancelled by the large radiative corrections, at the values of $Q$ available in the late 70's.
Ultimately the QCD explanation won out, because it was a theory not a model.

\begin{figure}[t]
\centering
\includegraphics[width=0.5\textwidth,angle=270]{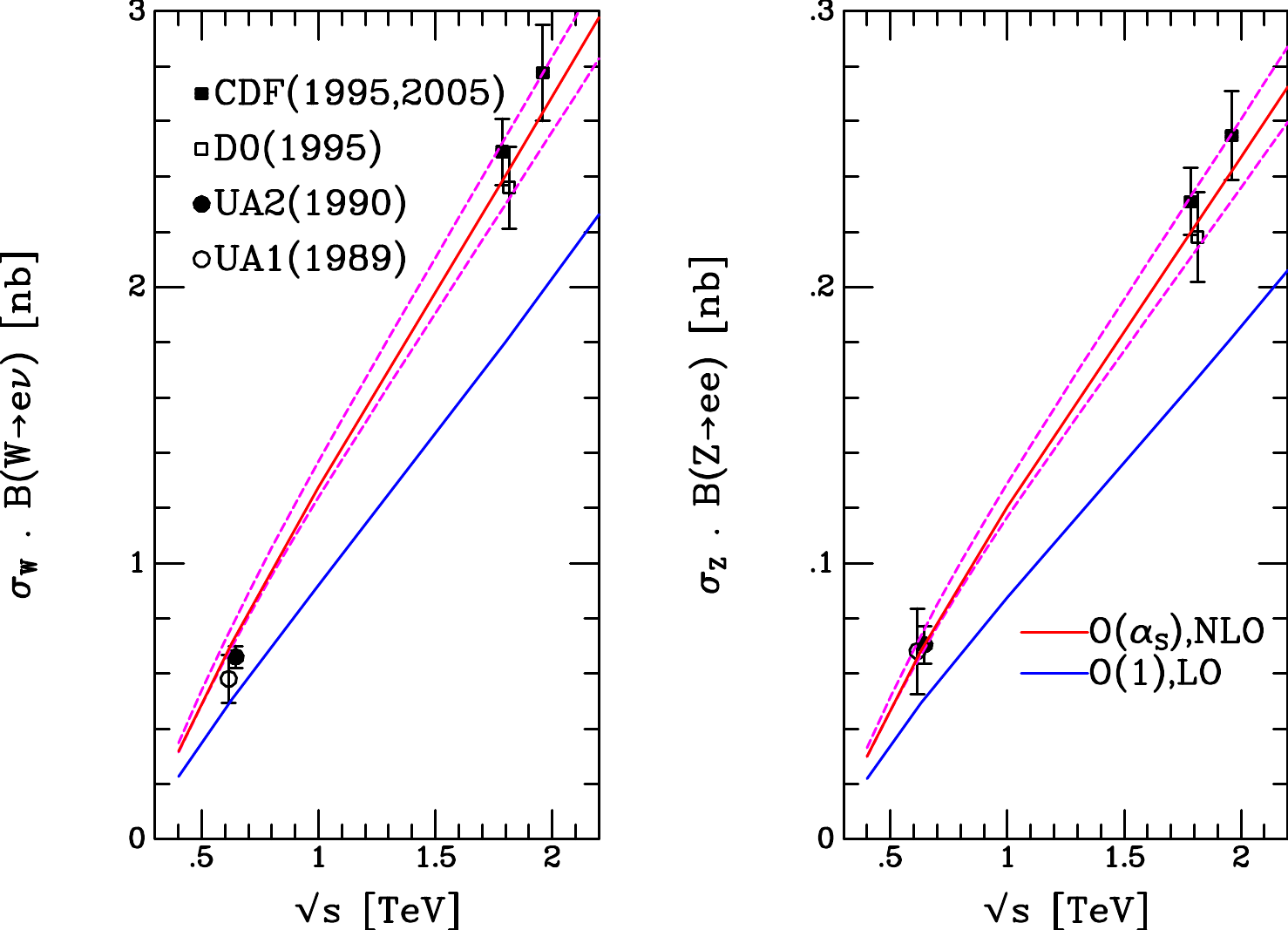}
\caption{LO and NLO predictions for $W$ and $Z$ production which proceed through a Drell-Yan type processes.
The pink dashed lines indicate theoretical uncertainties on the NLO prediction.}
\label{fig:wztot}
\end{figure}

Another source of uncertainty was the lack of knowledge of the antiquark distribution functions.
Lepton pair data from $\bar{p}p$ collisions was valuable in this regard, since it involved primarily
valence distributions~\cite{Saclay-CERN-CollegedeFrance-EcolePoly-Orsay:1979yqw}.
Fig.~\ref{fig:wztot} shows the influence of the NLO QCD corrections for $W$ and $Z$
production at the higher energies of the S$p\bar{p}$S and the Tevatron. The correction is still sizeable and needed for
agreement with the data.

\subsection{Factorization beyond leading logs}
\subsubsection{Deep Inelastic Scattering}
Deep inelastic scattering was first treated beyond leading logs using the operator product expansion. 
The two-loop anomalous dimensions were calculated in 1977 and 1978~\cite{Floratos:1977au,Floratos:1978ny} and
the one-loop process-dependent coefficient functions for DIS can be found in
refs.~\cite{Bardeen:1978yd,DeRujula:1976baf,Calvo:1977ba}.
Ref.~\cite{Bardeen:1978yd} made the important observation that inverse powers of $\epsilon$ also
appear with constant terms and proposed the modified minimal subtraction scheme ($\overline{MS}$)
in which renormalization constants absorb these constants as well as the pure pole pieces,
\beq
\frac{1}{\epsilon} \to \frac{1}{\epsilon}+\ln(4 \pi)-\gamma_E,\;\; \gamma_E = \text{Euler–Mascheroni~constant}.
\eeq
The phenomenological consequences of NNLO for DIS will be treated in other talks at this conference.

\subsubsection{Factorization for other hard processes}
Arguments that factorization should hold beyond leading logs were presented in 1978 in refs.~\cite{Ellis:1978sf,Ellis:1978ty,Amati:1978by}.
The simplest test case to assess these factorization ideas was the Drell-Yan process.
A systematic treatment of the Drell-Yan process was undertaken by Collins, Soper and Sterman~\cite{Collins:1982wa,Collins:1983ju}
taking special care to assess the role of soft gluons~\cite{Collins:1988ig}.
They argue that the proof of the factorization is as important as the proofs of renormalization and that the proofs are expected to be of
similar complexity. In further papers these authors extended the treatment to hadron-hadron scattering~\cite{Collins:1985ue}.
For an overall review of these efforts for factorization see ref.~\cite{Collins:1989gx}.

Using a formalism inspired by ref.~\cite{Ellis:1978sf,Ellis:1978ty},
Curci, Furmanski and Petronzio~\cite{Curci:1980uw,Furmanski:1980cm,Ellis:1996nn}.
calculated the splitting functions in order $O(\as^2)$ directly in $x$ space\footnote{These results
in $x$-space for the  spacelike case~\cite{Furmanski:1980cm} differ from those already obtained with the
operator product expansion formalism~\cite{Floratos:1978ny} in the part of $P_{gg}$ proportional to $C_A^2$, $(C_A=N=3)$.
The origin of the discrepancy in the $n$-space calculation was found in ref.~\cite{Hamberg:1991qt}.}. The $x$ space results
are much more compact than their moment-space predecessors.

\section{IR safety}
\label{IRsafety}
QCD predicts a colour octet of vector gluons, but little was experimentally known about these gluons. By the early 80's it was
known that about half of the momentum of the proton was carried by non-electrically charged particles. For example in
ref.~\cite{Abramowicz:1981be} the fraction of the proton momentum carried by gluons in the QCD theory was measured as $0.55\pm 0.11$.

In 1975 it was discovered at SPEAR that at energies above 5~GeV hadrons were produced in back-to-back jets~\cite{Hanson:1975fe}.
Ellis, Gaillard and Ross~\cite{Ellis:1976uc} made the suggestion that at higher energies broadening of these
jets should occur because of gluon radiation and eventually coplanar three-jet like
events should emerge, with differing probabilities for vector and scalar gluons.
However the sphericity variable which they used to describe the 3-jet structure
would not have given a perturbative prediction for the three jet cross section if
calculated beyond leading order. It was therefore not useful for precision QCD
which is the topic of this talk.

\begin{figure}[t]
\centering
\includegraphics[width=0.5\textwidth,angle=270]{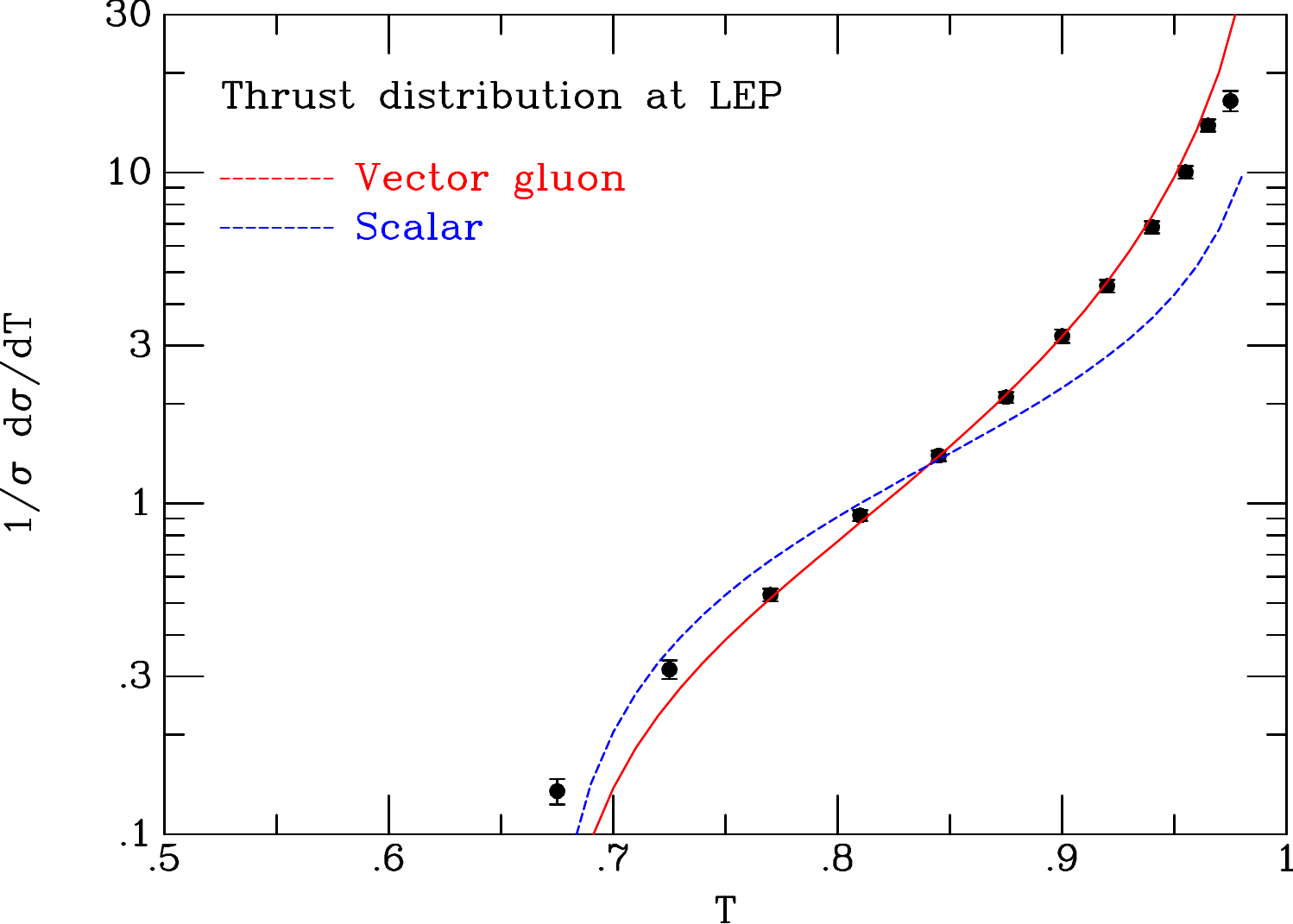}
\caption{Thrust distribution at LEP from the DELPHI collaboration~\cite{DELPHI:1992qrr}.}
\label{fig:thrust}
\end{figure}

This impasse was overcome in the
work of Sterman and Weinberg~\cite{Sterman:1977wj}, who posited that any quantity that made sense in the theory with massless quarks, or equivalently was insensitive to
the emission of soft or collinear partons, would be calculable purely in perturbation theory. Such quantities are said to be IR safe,
and more practical examples of IR safe variables 
were quickly proposed in refs.~\cite{Georgi:1977sf,Parisi:1978eg}. We shall focus here on the thrust~\cite{Farhi:1977sg}.
\beq
T=\max _{\vec{n}}\frac{\sum _{i}|\vec{p}_{i}\cdot \vec{n}|}{\sum _{i}|\vec{p}_{i}|}.
\eeq
$\vec{p}_{i}$ is the momentum of the $i$-th final-state particle. $\vec{n}$ is a unit vector called the "thrust axis".
Two jet events reside at $T=1$, whereas completely spherically symmetric events
have $T=0.5$.
Fig.~\ref{fig:thrust} shows the thrust distribution at LEP revealing that vector gluons are favoured over scalar gluons.

\section{Resummation}
\label{Resummation}
When the kinematics of real radiation is restricted additional, (double) logarithms are generated, because the real
radiation is forced to be soft and/or collinear. Virtual corrections are evidently unaffected. When these logarithms $L$
become large, $L \sim 1/\alpha_{s}$ perturbation theory breaks down and higher order terms are of the same order as leading order
terms, $(\as L)^n L \sim \as L^2$. Resumming large logarithms gives the correct physical Sudakov suppression allowing physical predictions
to be made in resummed perturbation theory. 

Double logarithms for suitable observables exponentiate exactly. Constraining the real radiation to be less than a scale, $k$ we
obtain the restricted real radiation rate, $\Sigma$,
\beq
\Sigma(k) = \int_0^k \;dk^\prime\; \frac{1}{\sigma_{\rm Born}} \;\frac{d \sigma}{d k^\prime} \sim \exp({ \as^n L^{n+1}
  +\as^n L^{n}
+\as^n L^{n-1}+\ldots})
\eeq
The first tower of terms
in the exponential is referred to as the leading log (LL) and is enhanced with respect to the Born cross section when $L \sim 1/\alpha{s}$,
the second tower of terms is the next-to-leading log (NLL) and is of the same order as the Born cross section 
and the third tower of terms is the (NNLL) and is the same order as NLO corrections in this limit.

\begin{figure}[t]
\centering
\includegraphics[width=0.4\textwidth]{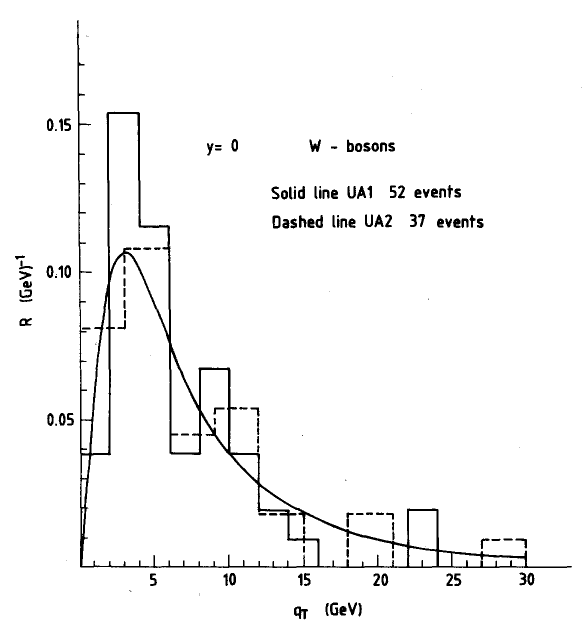}
\caption{$W$ boson $q_T$ distribution compared with data at the time of the $W$ boson discovery~\cite{Altarelli:1984pt}.}
\label{fig:Wbosons}
\end{figure}

Following an initial idea of Dokshitzer et
al.~\cite{Dokshitzer:1978yd}, this subject was first taken up by
Parisi and Petronzio~\cite{Parisi:1979se} for the case of muon pairs
produced in the Drell-Yan process. Working in the Fourier space
conjugate to the transverse momentum of the emitted boson to make the phase space constraint factorizable, they find
that for sufficiently large energy the full $k_T$ distribution is
calculable, including the region around $k_T=0$ which is dominated by the emission of semi-hard gluons of balancing $k_T$.
Subsequently Collins, Soper and Sterman~\cite{Collins:1984kg} set up a detailed formalism for the transverse momentum of vector bosons
produced in Drell-Yan type processes incorporating all the information about the leading and sub-leading coefficients known at the time.
In parallel, Altarelli et al.~\cite{Altarelli:1984kp,Altarelli:1984pt} provided phenomenological predictions. An historical example is shown in
Fig.~\ref{fig:Wbosons} for $W$-production which proceeds through a Drell-Yan like process.

In the new millenium the vision embodied in this section and the previous two was somewhat complicated by the discovery of
non-global logs~\cite{Dasgupta:2001sh}, extra logarithms which occur when observations are made in restricted regions of phase space. 
However this discovery falls outside the period of study in this review.

\section{Assembling physical cross sections}
To obtain distributions in NLO calculations requires combination of real radiation terms with $n+1$ partons
and virtual contributions with $n$ partons. The sum of the two terms is finite, by virtue of the KLN theorem.
However to obtain finite results for kinematic distributions new combination techniques are needed.
There are two techniques called subtraction and slicing.
We will use a simple model to describe the techniques.
Denote the real and virtual contributions as $I_R$ and $I_V$. The full result is given by $I=I_R+I_V$.
Let $x$ denote the energy
of the extra emitted parton in the real process, and $\epsilon$ a dimensional regularization parameter.
The real radiation contribution has an $F(x)/x$ emission spectrum, ($F(0)$ is finite),
whereas the virtual contribution, where there is no emitted
parton is independent of $x$, 
\beq
I_R = \int_0^1 \frac{dx}{x^{1+\ep}} F(x), \;\;I_V= \frac{1}{\ep} F(0) \, .
\eeq
\subsection{Subtraction}
The subtraction method~\cite{Ellis:1980wv,Kunszt:1989km,Frixione:1995ms,Catani:1996vz} proceeds by
subtracting a term that is simple enough to be integrated
and matches the singular piece of the real radiation,
\beqn
I &=& \int_0^1 \frac{dx}{x^{1+\ep}} [F(x)-F(0)]+\int_0^1 \frac{dx}{x^{1+\ep}} F(0) +I_V\, ,\nonumber\\
     &=& \int_0^1 \frac{dx}{x^{}} [F(x)-F(0)]-\frac{1}{\ep} F(0)+ \frac{1}{\ep} F(0)\, .
\eeqn
\subsection{Slicing}
The slicing method~\cite{Fabricius:1981sx,Giele:1991vf} proceeds by dividing the real contribution into resolved
and unresolved radiation by introducing a parameter $\delta$. 
In the unresolved radiation region the matrix element,
(in this case the function $F(x)$) is approximated by a simpler formula,
\beqn
I &=& \int_{\delta}^1 \frac{dx}{x^{1+\ep}} F(x)+ \int_0^{\delta} \frac{dx}{x^{1+\ep}} F(x) +I_V \, , \nonumber \\
& \approx & \int_{\delta}^1 \frac{dx}{x^{1+\ep}} F(x)+ \int_0^{\delta} \frac{dx}{x^{1+\ep}} F(0) +  \frac{1}{\ep} F(0)\, ,\nonumber \\
    &=& \int_{\delta}^1 \frac{dx}{x^{1+\ep}} F(x)  -\frac{1}{\ep}\delta^{-\ep} F(0)+  \frac{1}{\ep} F(0)\, , \nonumber \\
    &\approx & \int_{\delta}^1 \frac{dx}{x} F(x)  + F(0)\, \ln \delta \, .
\eeqn
The slicing method is simple to implement but
subject to large numerical cancellations as the limit $\delta \to 0$ is taken.

\section{Managing the expression swell -- Computer Algebra}
The calculation of higher order corrections needs to be performed analytically, at least in part, because of the
divergences, regulated by dimensional regularization. This normally leads to large analytical expressions manipulated
by the use of computer algebra.
As shown in Table~\ref{tab:Computer_Algebra} the computer algebra systems were first conceived in the late 1960's.
However they became widely used in the period of study 1976-2000 with increasing computer memory and the
availability of compact desktop systems. Table~\ref{tab:Computer_Algebra} is a partial list of computer algebra
systems. It only displays systems which have either originated in the high-energy community or have been extensively used
by high energy physics practitioners.
There is a sense in which QCD is the perfect application for computer algebra. The division of amplitudes into
non-gauge invariant Feynman diagrams, when summed with the correct signs (assured by the use of computers),
often leads to simple final results.
\begin{table}[h]
\begin{center}
\begin{tabular}{|l|l|l|}
    \hline
    1967 & Schoonschip & Martinus Veltman\cite{Veltman} \\
    1967 & Reduce  & Anthony C. Hearn  \\
    1976 & Ashmedai & M. J. Levine and R. Roskies\\
    1968-1982 & Macsyma & MIT Laboratory for Computer Science \\
    1982 & Maxima (open source) & William Schelter\\
    1988 & Mathematica 1.0 & Wolfram Research\\
    1982 & Maple &  Maplesoft\\
    1983 & Schoonschip (Motorola 68000 version) & Martinus Veltman and David Williams\\
    1989 & Form (open source 2010) & Jos Vermaseren \\
    1999 & Ginac & C. Bauer, A. Frink and R. Kreckel\\
    \hline
\end{tabular}
\caption{List of computer algebra systems commonly used in high-energy physics.}
\label{tab:Computer_Algebra}
\end{center}
\end{table}

For one-loop diagrams with five or fewer external legs, (which was the case for the the simplest
applications which were first calculated), the method of Passarino and Veltman~\cite{Passarino:1978jh}
gave an easily
programmable method of reducing one-loop calculations to scalar integrals, (integrals containing no
powers of the loop momenta in the numerator).
Results for all finite scalar integrals for boxes, triangles, bubbles and tadpoles
are given in ref.~\cite{tHooft:1978jhc}. These integrals were coded
in ref.~\cite{vanOldenborgh:1990yc,Hahn:1998yk}. 
Divergent scalar integrals were calculated as needed.
A complete set of references for divergent one-loop scalar integrals is available in ref.~\cite{Ellis:2007qk}.

Automatic procedures for the calculation of one loop matrix elements, require
\begin{itemize}
\item The generation of the Feynman diagrams
\item The manipulation of the algebraic expressions for the diagrams to compute the contribution to the
  matrix element squared.
\item Substitution of the numerical values for the scalar integrals to yield the numerical result.
\end{itemize}
As an example of these steps implemented mainly as a Mathematica computer code, we cite the suite of programs,
Feynarts~\cite{Kublbeck:1990xc}, Feyncalc~\cite{Mertig:1990an}, Formcalc~\cite{Hahn:1998yk} and Looptools~\cite{Hahn:1998yk}.

\begin{figure}[t]
\centering
\includegraphics[width=0.75\textwidth]{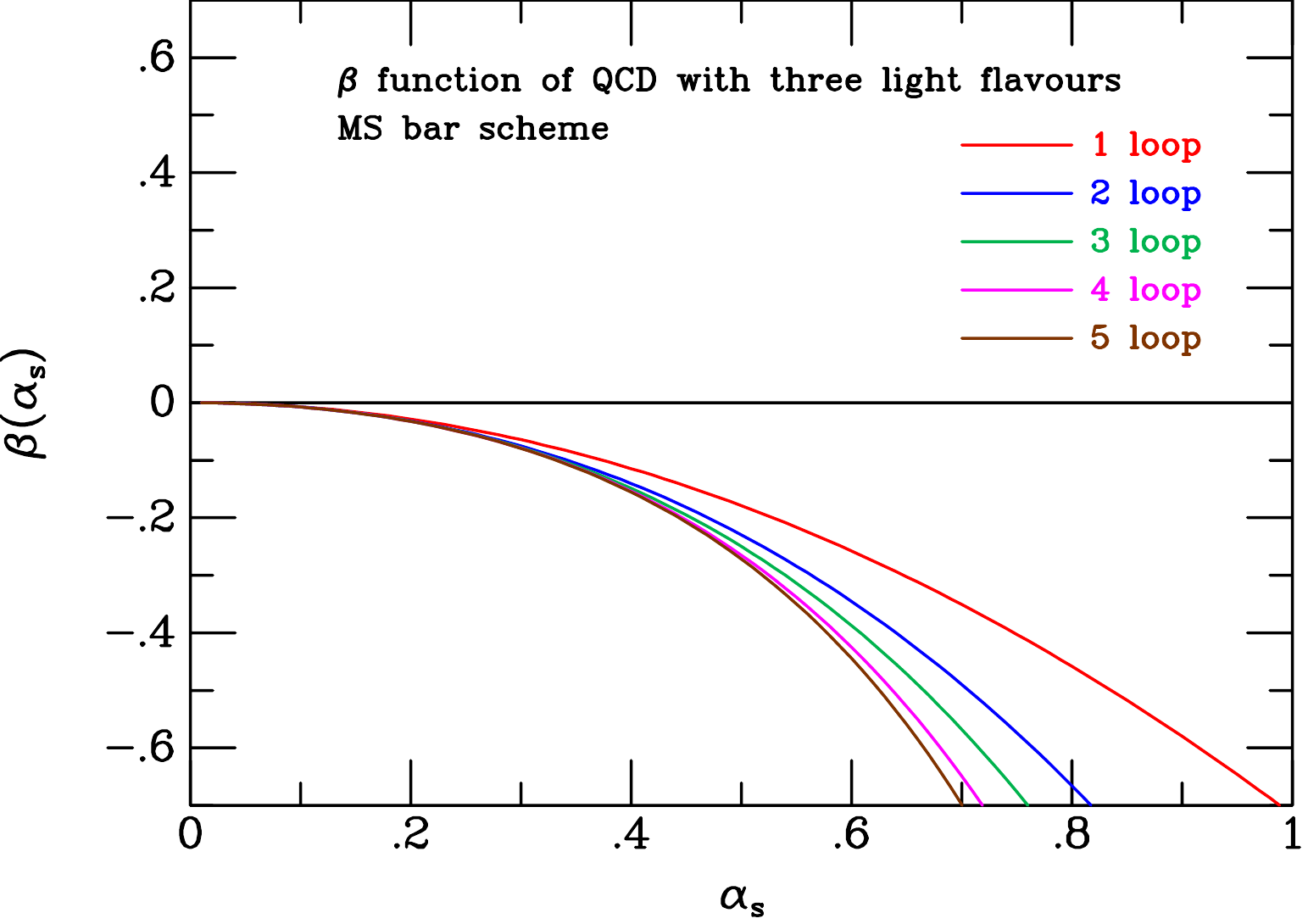}
\caption{Beta function in the $\overline{MS}$ scheme, including higher order
corrections~\cite{Caswell:1974gg,Jones:1974mm,Tarasov:1980au,Larin:1993tp,vanRitbergen:1997va,Baikov:2016tgj,Herzog:2017ohr,Herzog:2025ngd}.
Only the five loop result lies outside the time period of this review.}
\label{fig:beta}
\end{figure}

For diagrams with a small number of external legs $n=2,3$ and hence only one or two kinematic scales,
one can go to very high order. For more complex diagrams with more kinematic scales, one is limited
to leading order or next-to-leading order. As an example of variables of the first type, Fig.~\ref{fig:beta}
shows the 2025 results for the $\beta$-function as a function of $\as$.
For three loops and beyond these results required extensive use of computer algebra, especially
FORM~\cite{Vermaseren:1992vn,Vermaseren:2000nd} including custom enhancements.
Although the higher order terms in
the $\beta$-function are scheme-dependent these terms can become important in combination with other quantities,
such as the cusp anomalous dimension, in resummation calculations where one probes regions where the value of $\as$ is large.
\section{Avoiding the expression swell -- Spinor techniques}

Amplitudes involving massless particles can be expressed in terms of spinor products. Expressed in these terms
the structure of the amplitudes is manifest, in particular their form in the limit of collinear or soft momenta.
Consider an amplitude of the form $A_0 + \as A_1$. Traditional methods would calculate the lowest order amplitude squared, $|A_0|^2$
or the NLO matrix element squared $2 \as {\rm Re}\{A_0^{*} A_1\}$, leading to a proliferation of terms. In contrast direct calculation of the amplitude
expressed in spinor products can avoid the expression swell\footnote{Ref.~\cite{Schwartz:2014sze} gives an introduction to Weyl spinors.}.

Taking all particles to be outgoing, the massless spinors that we require are the wave functions associated with outgoing
fermions and anti-fermions.
In the Weyl representation these spinors satisfy the massless Dirac equation for fermions,
\begin{equation}
\bar{u}_\pm(p) \slsh{p}=0,\;{\rm where}~\bar{u}_-(p)=\big( \,\mathbf{0} \, , \,\langle p |^{\beta}\big),\;\;
\bar{u}_+(p)=\big(\, [ p |_\db \, ,\, \mathbf{0} \,\big)
\end{equation}
and anti-fermions
\begin{equation}
\slsh{p} v_\pm(p)=0,\;{\rm where}~v_-(p)=\left(\begin{matrix}
         \mathbf{0}  \\
         |p\rangle_{\alpha}\end{matrix} \right),\;\;
v_+(p)=\left( \begin{matrix}
     {|p]}^\da\\
    \mathbf{0}\end{matrix} \right)
\end{equation}
Thus for massless particles the four component spinors of the Dirac equation are not needed and
two component Weyl spinors give a complete description. Note that for massless particles, $v_{\pm}(p)=u_{\mp}(p)$.
$\alpha,\da$ are van der Waerden SL$(2,\mathbb{C})$ indices of the two-component Weyl spinors. 
In the QCD literature we dispense with the SL$(2,\mathbb{C})$ indices and write
the Lorentz-invariant spinor products as,
\beq
\langle p_i |^{\alpha} |p_j\rangle_{\alpha} \equiv \bar{u}_-(p_{i}) u_{+}(p_j) =\langle ij \rangle ,\;\;\;
[p_i |_{\db} |p_j]^{\db} \equiv \bar{u}_+(p_{i}) u_{-}(p_j)=[ij]\,.
\eeq
The spinor products are related to the dot-products of the massless momenta as follows,
\beq
\langle ij \rangle [ji] = 2 p_i \cdot p_j\, .
\eeq

The derivation of amplitudes in terms of spinor products was improved in steps.
The first step was the choice of polarization vector
made by the CALKUL collaboration~\cite{Berends:1981rb,Berends:1981uq,DeCausmaecker:1981jtq},
\begin{equation}
  \slsh{\varepsilon}(k)=\frac{1}{2\sqrt{2}}N
  \Big[\slsh{k} \slsh{q_+} \slsh{q_-}(1\pm \gamma_5)-
    \slsh{q_-} \slsh{q_+} 
    \slsh{k}(1\mp \gamma_5)\Big]\, .
\end{equation}
By appropriate choice of the arbitrary massless momenta $q_-$ and $q_+$ the calculation can be simplified.

A remarkable advance was made by researchers from Tsinghua University in China,
by simplifying the definition of polarization vector~\cite{Xu:1984qe,Xu:1985cs,Zhang:1985nk,Xu:1986xb}
so that it involved just one auxiliary massless momentum $n$ in addition to the momentum of the gluon, $k$,
\begin{equation} \label{Xuchoice}
  \varepsilon_{\mu}^{+}(k)
    =+\frac{\langle n|\gamma_\mu|k]}{\sqrt{2}\langle n k \rangle},\;\;\;
  \varepsilon_{\mu}^{-}(k)
    =- \frac{[n|\gamma_\mu|k\rangle}{\sqrt{2}[n k]} \, .
\end{equation}
Publication of these important ideas was unduly delayed\footnote{Ref.~\cite{Xu:1986xb}, a summary of Refs.~\cite{Xu:1984qe,Xu:1985cs,Zhang:1985nk},
took more than two years from receipt of the original manuscript in January
1985 to publication in early 1987.}
but Western authors were quick to utilize the method.

As an example of these ideas at work we first define colour-stripped amplitudes. Thus for the $n$-gluon tree amplitudes we have
\beq
-i {\cal A}^\tree_n = g^{n-2} \sum_{\sigma\in S_n/Z_n} 
\Tr(\si(1)\ldots\si(n)) A^\tree_n(\si(1),\ldots,\si(n)) \,, 
\label{TreeAmplitude}
\eeq
where $A_n^\tree$ are the colour-stripped partial amplitudes.
$\Tr(1\ldots n) \equiv \Tr(T^{a_1}\ldots T^{a_n})$,
with $a_i$ the colour index of the $i$-th external
gluon, and $S_n/Z_n$ is the set of non-cyclic
permutations of $\{1,2,\ldots,n\}$, corresponding to the set of
inequivalent traces. $T^{a_n}$ are the $SU(3)$ colour matrices in
the fundamental representation, normalized so that $\Tr(T^{a} T^{b})=\delta^{ab}$.

Certain classes of these colour stripped amplitudes vanish at tree level.
Using the polarization choice of Eq.~(\ref{Xuchoice}) it can be shown that
when fewer than two gluons have negative helicity and all the rest have positive helicity,
\beq
    {A}(1^{+},\cdots ,n^{+})=0,\;\;\;
    {A}( 1^+, \cdots i^-,\cdots,n^+ ) = 0.
\eeq
The first non-vanishing case occurs when two gluons have negative helicity. Such amplitudes are known as ``maximally helicity violating'' and have an extremely simple form in terms of spinor products, independent of the number of gluons present,
\beq \label{Parke-Taylor}
{A}(1^{+}\cdots i^{-}\cdots j^{-}\cdots n^{+})={\frac {\langle i\;j\rangle ^{4}}{\langle 1\;2\rangle \langle 2\;3\rangle \cdots \langle (n-1)\;n\rangle \langle n\;1\rangle }}
\eeq
This formula was first suggested by Parke and Taylor~\cite{Parke:1986gb} and later proved by induction by Berends and Giele~\cite{Berends:1987me}. 
Note how remarkable it is that at large $n$ the net result of many Feynman diagrams is this simple formula.

In the 21st century, formulae such as Eq.~(\ref{Parke-Taylor}) have given rise to a whole subfield of amplitudes~\cite{Elvang:2013cua}.
In this subfield amplitudes which transform properly under the little group are the primary building blocks
and quantum fields involving off-shell particles are never introduced\footnote{The little group
is the group of transformations that leave the momentum of an on-shell particle invariant.}.

\section{Key applications at NLO and beyond}
\subsection{NLO}
With the tools of sections~\ref{factorization}-\ref{Resummation}
in place, it was now time to actually calculate physical results, using the
tools, either in combination or separately. 
A complete review of the large number of NLO calculations in the period 1976-2000 is certainly not possible in this presentation.
Instead I shall review a limited number of important calculations, and emphasize the new technical features.

\subsubsection{3-jet production in $e^+ e^-$ annihilation.}
\begin{figure}[t]
\centering
\includegraphics[width=0.5\textwidth]{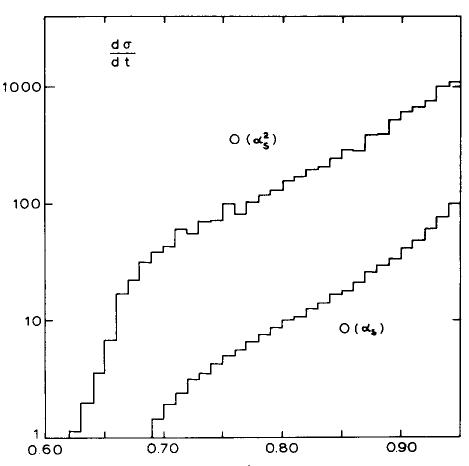}
\caption{The scaling factors $(1/\sigma_0)(\pi/\as)$ and
  $(1/\sigma_0)(\pi/\as)^2$ have been used for the $O(\as)$ and $O(\as^2)$ histograms~\cite{Vermaseren:1980qz}.
The relative size of the two contributions can be estimated at $Q \sim 30$~GeV by setting $\as/\pi \sim 0.06$.}
\label{fig:Vermaserenthrust}
\end{figure}
The $O(\as^2)$ corrections to jet production in $e^+ e^-$ collisions
were first calculated~\cite{Ellis:1980wv} for the $C$-parameter, an infra-red safe global shape variable derived
from the super inclusive cross sections of ref.~\cite{Parisi:1978eg}. For $Q=30$~GeV, $\as=0.188$ it is found that the corrections 
are so large that they cast doubt on the applicability of perturbation theory.
As shown in Fig.~\ref{fig:Vermaserenthrust} this conclusion is confirmed also for the
thrust variable~\cite{Vermaseren:1980qz,Kunszt:1980vt}.

\subsubsection{4-jet production in $e^+ e^-$ annihilation.}
In $e^+ e^-$ annihilation, four jet events calculated at leading order presents the first instance where
the three-gluon vertex plays a role. Refs.~\cite{Nachtmann:1982xr,Bengtsson:1988qg} construct angles using the
momenta of the four jets to make the influence of the three-gluon vertex manifest.
Experiment favours QCD over Abelian models in which the three-gluon vertex is absent.

One-loop amplitudes for $e^+ e^-$ to four partons were calculated in ref.~\cite{Bern:1997sc}. These formula are also
useful for the crossed processes, $q\qb \to W/Z+2$~jets and for three-jet production in DIS.
The rates for electron-positron annihilation into four jets at next-to-leading order in $\as$ are given~\cite{Dixon:1997th}.
For the Durham jet algorithm~\cite{Catani:1991hj}, the NLO corrections to the overall rate are large and significant
renormalization-scale dependence remains in the NLO prediction.

\subsubsection{Parton distribution functions}
Predictions for high energy processes involving hadrons depend crucially on knowledge of 
the parton distribution functions. Early efforts to provide parton distributions consistent
with the leading order DGLAP equation based on simple asymptotic forms with $Q^2$-dependent 
exponents were given in \cite{Buras:1977yj,Duke:1983gd}. A more industrial approach to global
fits to data consistent with the NLO DGLAP equation started in the late 80's and early 90's~\cite{Martin:1987vw,Botts:1993shg}.
These fits were based primarily on lepton-nucleon scattering, although some high-energy collider data was used.
The challenges in these fits were corrections for nuclear effects,
the strong correlations between the shape of the gluon and the fitted value of $\as$, and at times
inconsistent data sets~\cite{Martin:1988nk}.
In ref.~\cite{Gluck:1991ng} an additional theoretical assumption that the gluon and sea quark distributions were completely
generated by radiation from a valence quark distribution fixed at low scale was made. This approach was found to be too restrictive to
accommodate more detailed data.
The addition of $ep$ data from HERA reduced the need for nuclear corrections
and gave much better control of the gluon distribution~\cite{Martin:1996as}.
A review of the status of parton distribution functions in 1992 is provided by the article of Owens and Tung~\cite{Owens:1992hd}.

\subsubsection{Two jet production in hadron-hadron collisions}
\begin{figure}[t]
\centering
\includegraphics[width=0.5\textwidth,angle=270]{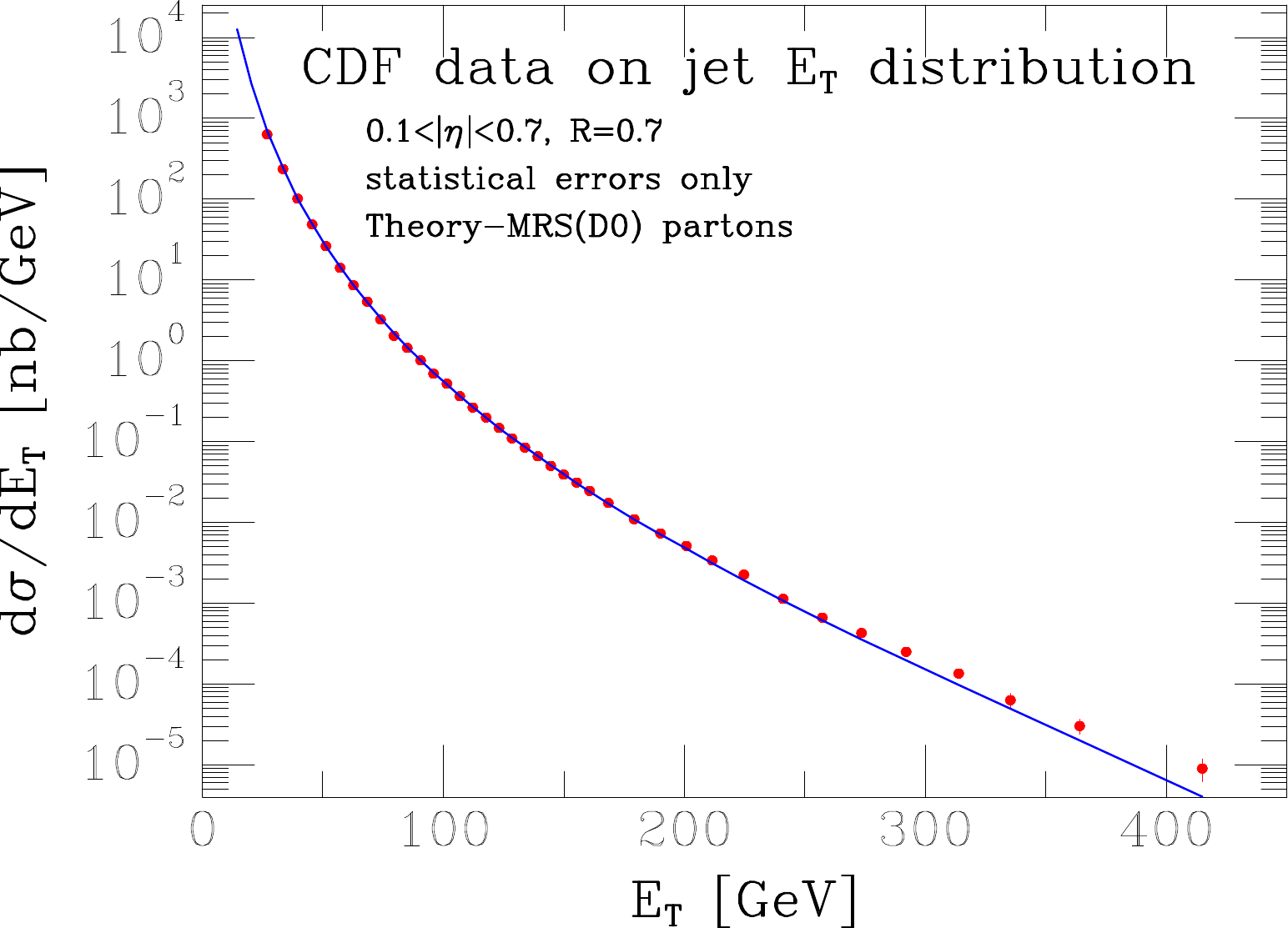}
\caption{Jet $E_T$ distribution from the CDF
collaboration~\cite{CDF:1996yow}, compared
with a next-to-leading order QCD prediction~\cite{Ellis:1990ek}.}
\label{fig:anwar}
\end{figure}

The lowest order $2 \to 2$ processes were calculated in ref.~\cite{Combridge:1977dm}.
The NLO contribution to the matrix element squared for all parton subprocesses were calculated in 1985~\cite{Ellis:1985er}.
These results were assembled into a prediction for the NLO jet rate in ref.~\cite{Ellis:1990ek} as shown in Fig.~\ref{fig:anwar}.
Simpler results for the two-to-two amplitudes were later obtained using spinor methods~\cite{Bern:1990cu,Kunszt:1993sd}.

\subsubsection{Three jet production in hadron-hadron collisions}
Spinor methods allowed the calculation of one-loop corrections to five gluon amplitudes~\cite{Bern:1993mq} in 1993,
followed shortly thereafter by the one-loop radiative corrections to the helicity amplitudes of
QCD processes involving four quarks and one gluon~\cite{Kunszt:1994nq} and the 
one-loop corrections to two quark three gluon amplitudes~\cite{Bern:1994fz}. These results show the superiority of the spinor methods over the older
methods used in ref.~\cite{Ellis:1985er}.
However the assembly of these results into a full prediction of 
next-to-leading order calculation of three jet observables in hadron-hadron collisions was not available
until 2003~\cite{Nagy:2003tz}.

\subsubsection{Top production}
\begin{figure}[t]
\centering
\includegraphics[width=0.4\textwidth,angle=270]{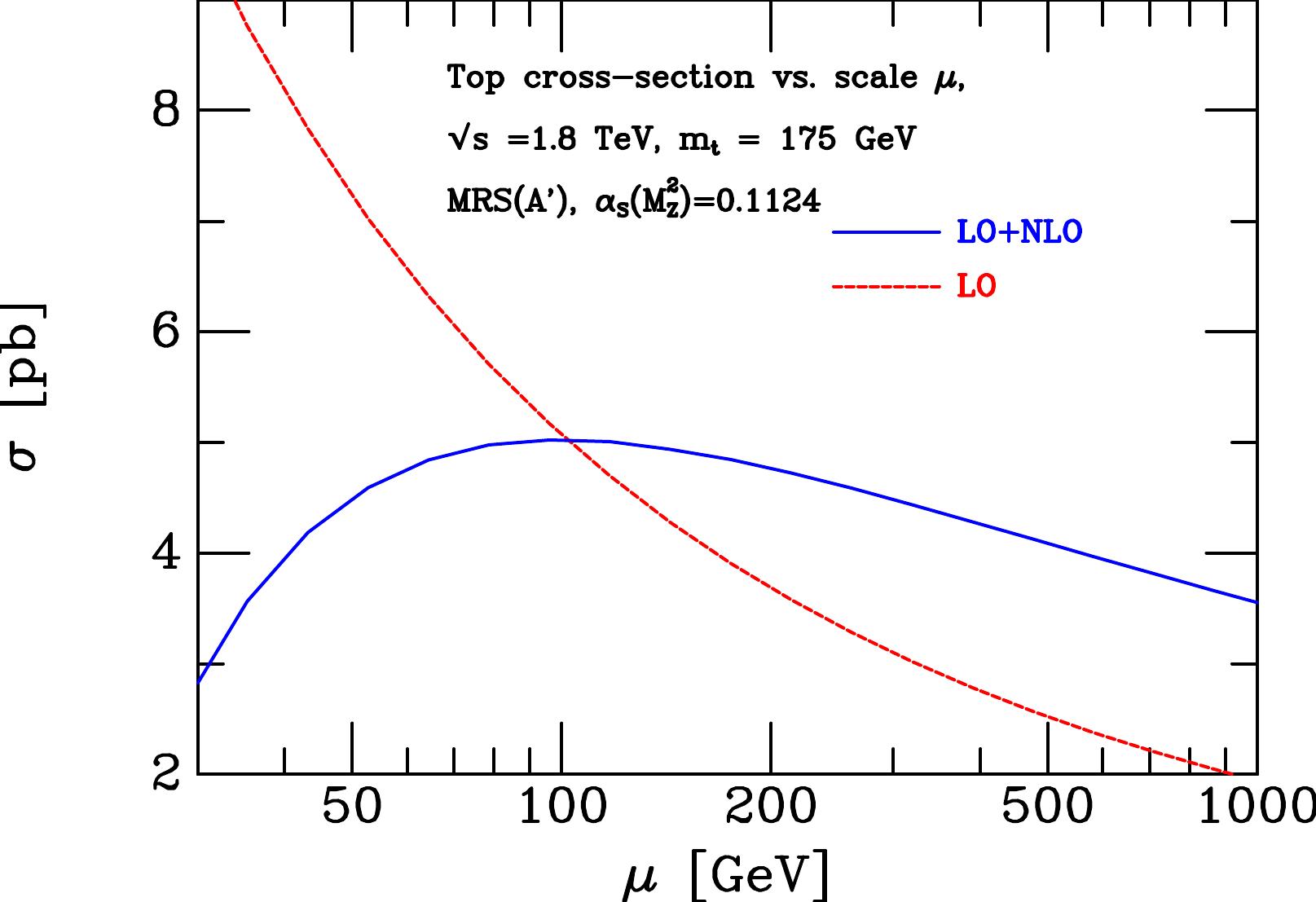}
\caption{Scale dependence in LO and NLO for top production.}
\label{fig:mu_175_1800}
\end{figure}

The production of heavy quarks with a mass much larger than the scale of
the strong interactions can be calculated in perturbative QCD~\cite{Collins:1985gm}.
NLO results were obtained in refs.~\cite{Nason:1987xz,Nason:1989zy} by Nason et al.~and subsequently 
by Beenakker et al.~\cite{Beenakker:1988bq,Beenakker:1990maa}. Fig.~\ref{fig:mu_175_1800} shows 
the reduced dependence on the scale choice $\mu$ for top production in $p \bar{p}$ collisions when NLO corrections are included.

In addition to accurate values for the top production cross-section, perturbative QCD played a significant role in
the discovery of the top quark by determining the backgrounds. These backgrounds were $W$ and $Z$-production associated with two or more jets. The  
VECBOS parton-level Monte Carlo~\cite{Berends:1990ax} modelled these processes at leading order in the QCD coupling using early implementations of
spinor techniques. NLO predictions for W+2~jet processes were only obtained early in the 21st century~\cite{Campbell:2002tg},
recycling NLO virtual matrix elements for
$e^+ e^-$+~{four jets}~\cite{Bern:1997sc}. On the other hand, QCD corrections to $W$-boson production in association with a pair of $b$-quarks
were already available before the turn of the millenium~\cite{Ellis:1998fv}.
\subsubsection{Vector boson pair production}
Ref.~\cite{Dixon:1998py} gave compact one-loop QCD corrections to the helicity amplitudes for the processes
$q\qb \to W^+ W^-, Z Z, W^\pm Z, W^\pm \gamma$, or $Z\gamma$,
including the subsequent decay of each massive vector boson into a pair of leptons.
These processes are important as fundamental tests of the electroweak standard model at high energy,
and as backgrounds in the search for the Higgs boson.
Phenomenological results~\cite{Dixon:1999di,Campbell:1999ah} show that inclusion of NLO leads to about a
$30\%$ enhancement both for $p\bar{p}$ at $\sqrt{s}=2$~TeV and $pp$ at $\sqrt{s}=14$~TeV.

\subsection{NNLO and beyond}
An important early result was the calculation of the corrections to the Drell-Yan process
at NNLO~\cite{Matsuura:1988sm,vanNeerven:1991gh,vanNeerven:1999ca} in 1988-1991. Recall that the K-factor
for the $\as$ corrections was large. At low values of $Q$ that called into question
the stability of the perturbative series; for massive vector boson production
at higher values of $Q\approx 80-90$~GeV the corrections are still of order $30\%$.
The results at NNLO show that the additional correction of order $\as^2$ is not large and 
drastically reduces the scale dependence, see Matsuura and Plothow-Besch in ref.~\cite{Jarlskog:1990bk}.

An important step in standardizing the calculation of massless two loop amplitudes
was the publishing of a general formula for the structure of singular terms~\cite{Catani:1998bh}.
Around the turn of the millenium the first two loop amplitudes were calculated using spinor methods~\cite{Bern:2000dn}.

\begin{figure}[t]
\centering
\includegraphics[width=0.35\textwidth,angle=270]{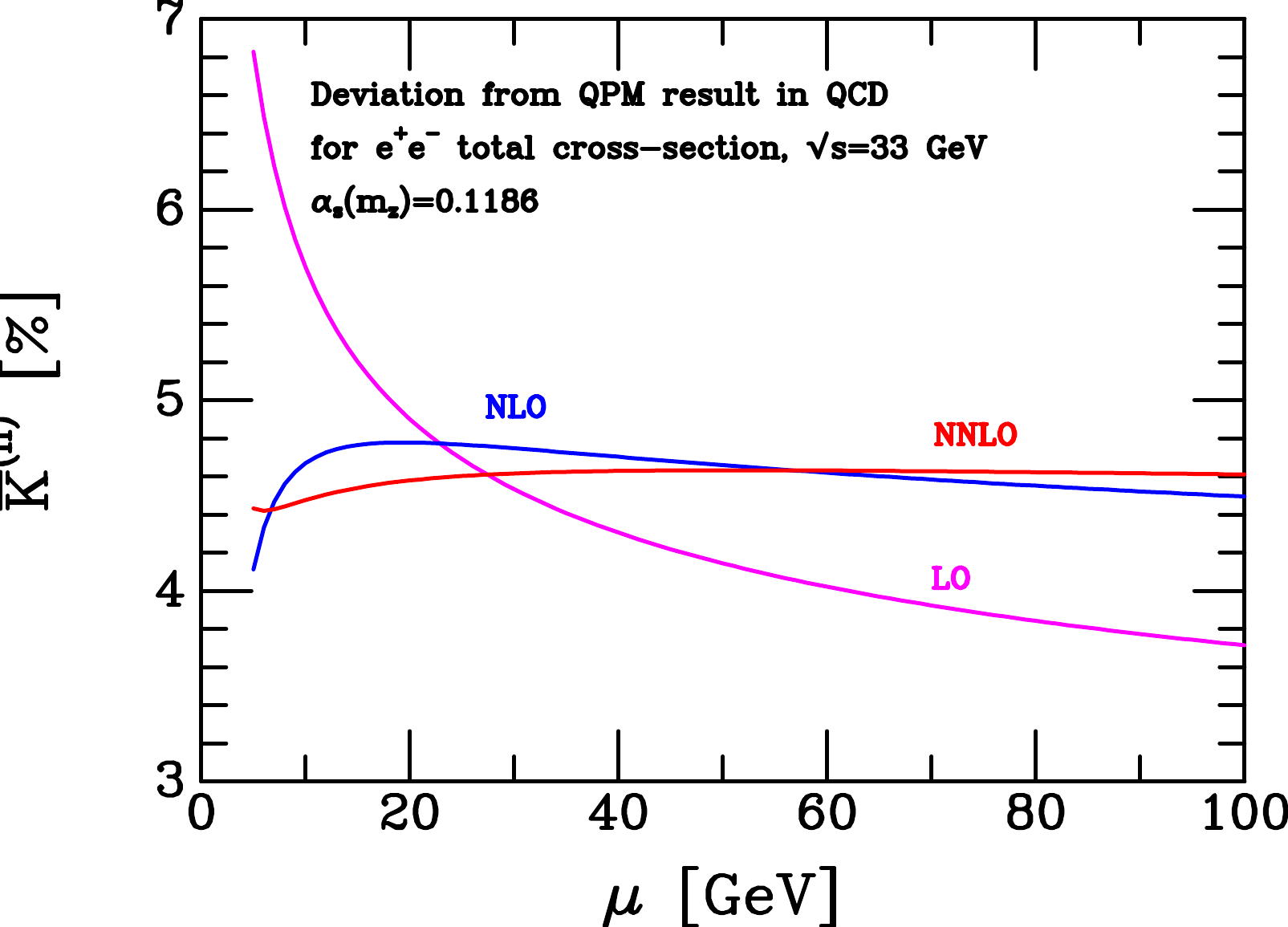}
\caption{QCD corrections to the total $e^+ e^-$ cross section as a percentage.
The quantity $\overline{K}^{(n)} = K^{(n)} -1$,
where $K^{(n)}$ denotes the QCD prediction for $K_{QCD}$ truncated at 
$O(\as^n)$, as a function of the renormalization scale $\mu$~\cite{Surguladze:1990tg,SurguladzeErratum,Gorishnii:1990vf}.}
\label{fig:mudep}
\end{figure}

The total cross section for $e^+ e^-$ collisions has been now calculated up to $O(\as^4)$~\cite{Baikov:2012zn}.
The impact of the first three terms
as a function of the scale is shown in Fig.~\ref{fig:mudep}.

\section{Technical advances}
I would like to briefly point out that a number of software and hardware
advances greatly facilitated the scientific communication and collaboration,
not just for precision QCD, but also for all of mathematical science.
\begin{table}[h]
\begin{center}
\begin{tabular}{|l|l|l|}
    \hline
    Date & Technical advance & Originator \\
    \hline
    1975 & DECnet network protocols  & Digital Equipment Corporation \\
    1981   & BITNET network & Ira H. Fuchs and Greydon Freeman \\
    \hline
    1976   & IBM 3800 laser printer & IBM\\
    1977   & Xerox9700 laser printer & Xerox corporation \\
    1978   & Versatec 8272 plotter & Xerox corporation \\
    1980   & HP 2680A laser printer & Hewlett Packard \\
    \hline
    1978   & \TeX  & Donald E. Knuth \\
    1985   & \LaTeX~2.09 & Leslie Lamport \\
    1993   & Portable Document Format (PDF) & Adobe\\
    \hline
    1991   & arXiv & Paul Ginsparg\\
    \hline
\end{tabular}
\end{center}
\caption{Technical advances in the period 1976-2000}
\label{Technical_advances}
\end{table}
These advances and their dates of arrival are listed in Table~\ref{Technical_advances}.
These advances are so important,
it is sometimes hard to imagine the world without them. 
The influence 
of the world-wide web will be discussed elsewhere in this conference.

\section{Epilogue}
The period 1976-2000 established many techniques for precise calculation in QCD.
The new millenium has given further astonishing advances in precision QCD. In synthesis 
these advances are as follows. For most practical purposes the problem of NLO QCD has been completely solved.
Processes with a small number of legs (2 or 3) can be calculated to
very high order, whereas currently only some processes with a greater number of legs are known
at two-loop order (or rarely three-loop order). These calculations are always performed using computer algebra,
oftentimes using codes written by others and shared amongst the community.

But I sometimes look back with nostalgia to a simpler time when pencil, eraser
and paper calculations were the modus operandi. To be sure it was laborious, but the
process was totally absorbing and the thrill of arriving at the simple final answer
was glorious.
\acknowledgments
I acknowledge receipt of a Leverhulme Emeritus Fellowship from the
Leverhulme Trust.

%% file: History_rkellis.bbl
\begin{thebibliography}{100}

\bibitem{Gross:1973zrg}
D.~J. Gross and Frank Wilczek.
\newblock {Asymptotically Free Gauge Theories. 2.}
\newblock {\em Phys. Rev. D}, 9:980--993, 1974.
\newblock \href {https://doi.org/10.1103/PhysRevD.9.980}
  {\path{doi:10.1103/PhysRevD.9.980}}.

\bibitem{Georgi:1974wnj}
Howard Georgi and H.~David Politzer.
\newblock {Electroproduction scaling in an asymptotically free theory of strong
  interactions}.
\newblock {\em Phys. Rev. D}, 9:416--420, 1974.
\newblock \href {https://doi.org/10.1103/PhysRevD.9.416}
  {\path{doi:10.1103/PhysRevD.9.416}}.

\bibitem{Wilson:1970ag}
Kenneth~G. Wilson.
\newblock {The Renormalization Group and Strong Interactions}.
\newblock {\em Phys. Rev. D}, 3:1818, 1971.
\newblock \href {https://doi.org/10.1103/PhysRevD.3.1818}
  {\path{doi:10.1103/PhysRevD.3.1818}}.

\bibitem{Gaillard:1974nj}
M.~K. Gaillard and Benjamin~W. Lee.
\newblock {$\Delta$ I = 1/2 Rule for Nonleptonic Decays in Asymptotically Free
  Field Theories}.
\newblock {\em Phys. Rev. Lett.}, 33:108, 1974.
\newblock \href {https://doi.org/10.1103/PhysRevLett.33.108}
  {\path{doi:10.1103/PhysRevLett.33.108}}.

\bibitem{Altarelli:1974exa}
Guido Altarelli and L.~Maiani.
\newblock {Octet Enhancement of Nonleptonic Weak Interactions in Asymptotically
  Free Gauge Theories}.
\newblock {\em Phys. Lett. B}, 52:351--354, 1974.
\newblock \href {https://doi.org/10.1016/0370-2693(74)90060-4}
  {\path{doi:10.1016/0370-2693(74)90060-4}}.

\bibitem{Appelquist:1974zd}
Thomas Appelquist and H.~David Politzer.
\newblock {Orthocharmonium and $e^+ e^-$ Annihilation}.
\newblock {\em Phys. Rev. Lett.}, 34:43, 1975.
\newblock \href {https://doi.org/10.1103/PhysRevLett.34.43}
  {\path{doi:10.1103/PhysRevLett.34.43}}.

\bibitem{Hoddeson:1997hk}
Lillian~H. Hoddeson, L.~Brown, M.~Riordan, and M.~Dresden, editors.
\newblock {\em {The Rise of the standard model: Particle physics in the 1960s
  and 1970s. Proceedings, Conference, Stanford, USA, June 24-27, 1992}}, 1997.
\newblock \href {https://doi.org/10.1017/CBO9780511471094}
  {\path{doi:10.1017/CBO9780511471094}}.

\bibitem{mythesis}
R.~Keith Ellis.
\newblock {\em Parity Violating Effects in Nuclear Processes}.
\newblock PhD thesis, February 1975.

\bibitem{Altarelli:1974ni}
Guido Altarelli, R.~Keith Ellis, L.~Maiani, and R.~Petronzio.
\newblock {The Structure of Parity Violating Strangeness Conserving Weak
  Nonleptonic Amplitudes in an Asymptotically Free Theory}.
\newblock {\em Nucl. Phys. B}, 88:215--236, 1975.
\newblock \href {https://doi.org/10.1016/0550-3213(75)90277-1}
  {\path{doi:10.1016/0550-3213(75)90277-1}}.

\bibitem{Marciano:1977su}
William~J. Marciano and Heinz Pagels.
\newblock {Quantum Chromodynamics: A Review}.
\newblock {\em Phys. Rept.}, 36:137, 1978.
\newblock \href {https://doi.org/10.1016/0370-1573(78)90208-9}
  {\path{doi:10.1016/0370-1573(78)90208-9}}.

\bibitem{Field:1976ve}
R.~D. Field and R.~P. Feynman.
\newblock {Quark Elastic Scattering as a Source of High Transverse Momentum
  Mesons}.
\newblock {\em Phys. Rev. D}, 15:2590--2616, 1977.
\newblock \href {https://doi.org/10.1103/PhysRevD.15.2590}
  {\path{doi:10.1103/PhysRevD.15.2590}}.

\bibitem{Combridge:1977dm}
B.~L. Combridge, J.~Kripfganz, and J.~Ranft.
\newblock {Hadron Production at Large Transverse Momentum and QCD}.
\newblock {\em Phys. Lett. B}, 70:234, 1977.
\newblock \href {https://doi.org/10.1016/0370-2693(77)90528-7}
  {\path{doi:10.1016/0370-2693(77)90528-7}}.

\bibitem{Feynman:1978dt}
R.~P. Feynman, R.~D. Field, and G.~C. Fox.
\newblock {A Quantum Chromodynamic Approach for the Large Transverse Momentum
  Production of Particles and Jets}.
\newblock {\em Phys. Rev. D}, 18:3320, 1978.
\newblock \href {https://doi.org/10.1103/PhysRevD.18.3320}
  {\path{doi:10.1103/PhysRevD.18.3320}}.

\bibitem{Chang:1970jk}
Shau-Jin Chang and P.~M. Fishbane.
\newblock {High-energy deep-inelastic electron-nucleon scattering in a
  renormalizable theory}.
\newblock {\em Phys. Rev. D}, 2:1084--1103, 1970.
\newblock \href {https://doi.org/10.1103/PhysRevD.2.1084}
  {\path{doi:10.1103/PhysRevD.2.1084}}.

\bibitem{Fishbane:1971mnh}
P.~M. Fishbane and J.~D. Sullivan.
\newblock {Inelastic e-p scattering in massive quantum electrodynamics}.
\newblock {\em Phys. Rev. D}, 4:2516--2533, 1971.
\newblock [Erratum: Phys.Rev.D 6, 1813--1813 (1972)].
\newblock \href {https://doi.org/10.1103/PhysRevD.4.2516}
  {\path{doi:10.1103/PhysRevD.4.2516}}.

\bibitem{Gribov:1971zn}
V.~N. Gribov and L.~N. Lipatov.
\newblock {Deep inelastic electron scattering in perturbation theory}.
\newblock {\em Phys. Lett. B}, 37:78--80, 1971.
\newblock \href {https://doi.org/10.1016/0370-2693(71)90576-4}
  {\path{doi:10.1016/0370-2693(71)90576-4}}.

\bibitem{Gribov:1972ri}
V.~N. Gribov and L.~N. Lipatov.
\newblock {Deep inelastic e p scattering in perturbation theory}.
\newblock {\em Sov. J. Nucl. Phys.}, 15:438--450, 1972.

\bibitem{Gribov:1972rt}
V.~N. Gribov and L.~N. Lipatov.
\newblock {$e^+e^-$ pair annihilation and deep inelastic e p scattering in
  perturbation theory}.
\newblock {\em Sov. J. Nucl. Phys.}, 15:675--684, 1972.

\bibitem{Christ:1972ms}
Norman~H. Christ, B.~Hasslacher, and Alfred~H. Mueller.
\newblock {Light cone behavior of perturbation theory}.
\newblock {\em Phys. Rev. D}, 6:3543, 1972.
\newblock \href {https://doi.org/10.1103/PhysRevD.6.3543}
  {\path{doi:10.1103/PhysRevD.6.3543}}.

\bibitem{Dokshitzer:1977sg}
Yuri~L. Dokshitzer.
\newblock {Calculation of the Structure Functions for Deep Inelastic Scattering
  and $e^+e^-$ Annihilation by Perturbation Theory in Quantum Chromodynamics,
  available at: \url{http://www.jetp.ras.ru/cgi-bin/dn/e\_046\_04\_641.pdf}}.
\newblock {\em Sov. Phys. JETP}, 46:641--653, 1977.

\bibitem{Parisi:1976qj}
G.~Parisi.
\newblock {An Introduction to Scaling Violations, available at
  \url{https://lib-extopc.kek.jp/preprints/PDF/1976/7611/7611077.pdf}}.
\newblock In {\em {11th Rencontres de Moriond}: {weak interactions and neutrino
  physics}}, pages 83--114, 4 1976.

\bibitem{Altarelli:1977zs}
Guido Altarelli and G.~Parisi.
\newblock {Asymptotic Freedom in Parton Language}.
\newblock {\em Nucl. Phys. B}, 126:298--318, 1977.
\newblock \href {https://doi.org/10.1016/0550-3213(77)90384-4}
  {\path{doi:10.1016/0550-3213(77)90384-4}}.

\bibitem{Cabibbo:1974qb}
N.~Cabibbo and M.~Rocca.
\newblock {The rho Bremsstrahlung,
  \url{https://inspirehep.net/files/b14c7cca9f17c4a8a969987aeb5de660}}.
\newblock 5 1974.

\bibitem{Baier:1973ms}
V.~N. Baier, Victor~S. Fadin, and Valery~A. Khoze.
\newblock {Quasireal electron method in high-energy quantum electrodynamics}.
\newblock {\em Nucl. Phys. B}, 65:381--396, 1973.
\newblock \href {https://doi.org/10.1016/0550-3213(73)90291-5}
  {\path{doi:10.1016/0550-3213(73)90291-5}}.

\bibitem{Parisi:2025nob}
Giorgio Parisi.
\newblock {From Bjorken Scaling to Scaling Violations}.
\newblock 6 2025.
\newblock \href {https://arxiv.org/abs/2506.03383} {\path{arXiv:2506.03383}}.

\bibitem{Feynman1972}
Richard~P Feynman.
\newblock {\em Photon-Hadron Interactions}.
\newblock CRC, 1972.
\newblock \href {https://doi.org/{https://doi.org/10.1201/9780429493331}}
  {\path{doi:{https://doi.org/10.1201/9780429493331}}}.

\bibitem{Christenson:1970um}
J.~H. Christenson, G.~S. Hicks, L.~M. Lederman, P.~J. Limon, B.~G. Pope, and
  E.~Zavattini.
\newblock {Observation of massive muon pairs in hadron collisions}.
\newblock {\em Phys. Rev. Lett.}, 25:1523--1526, 1970.
\newblock \href {https://doi.org/10.1103/PhysRevLett.25.1523}
  {\path{doi:10.1103/PhysRevLett.25.1523}}.

\bibitem{Lederman:1971pt}
L.~M. Lederman and B.~G. Pope.
\newblock {Production of intermediate bosons in strong interactions}.
\newblock {\em Phys. Rev. Lett.}, 27:765--768, 1971.
\newblock \href {https://doi.org/10.1103/PhysRevLett.27.765}
  {\path{doi:10.1103/PhysRevLett.27.765}}.

\bibitem{Drell:1970wh}
S.~D. Drell and Tung-Mow Yan.
\newblock {Massive Lepton Pair Production in Hadron-Hadron Collisions at
  High-Energies}.
\newblock {\em Phys. Rev. Lett.}, 25:316--320, 1970.
\newblock [Erratum: Phys.Rev.Lett. 25, 902 (1970)].
\newblock \href {https://doi.org/10.1103/PhysRevLett.25.316}
  {\path{doi:10.1103/PhysRevLett.25.316}}.

\bibitem{Politzer:1977fi}
H.~David Politzer.
\newblock {Gluon Corrections to Drell-Yan Processes}.
\newblock {\em Nucl. Phys. B}, 129:301--318, 1977.
\newblock \href {https://doi.org/10.1016/0550-3213(77)90197-3}
  {\path{doi:10.1016/0550-3213(77)90197-3}}.

\bibitem{Sachrajda:1977mb}
Christopher~T. Sachrajda.
\newblock {Lepton Pair Production and the Drell-Yan Formula in QCD}.
\newblock {\em Phys. Lett. B}, 73:185--188, 1978.
\newblock \href {https://doi.org/10.1016/0370-2693(78)90831-6}
  {\path{doi:10.1016/0370-2693(78)90831-6}}.

\bibitem{Humpert:1979hb}
B.~Humpert and W.~L. van Neerven.
\newblock {How to Regularize the Infrared and Mass Singularities in {QCD}}.
\newblock {\em Phys. Lett. B}, 89:69--75, 1979.
\newblock \href {https://doi.org/10.1016/0370-2693(79)90078-9}
  {\path{doi:10.1016/0370-2693(79)90078-9}}.

\bibitem{tHooft:1972tcz}
Gerard 't~Hooft and M.~J.~G. Veltman.
\newblock {Regularization and Renormalization of Gauge Fields}.
\newblock {\em Nucl. Phys. B}, 44:189--213, 1972.
\newblock \href {https://doi.org/10.1016/0550-3213(72)90279-9}
  {\path{doi:10.1016/0550-3213(72)90279-9}}.

\bibitem{Bollini:1972ui}
C.~G. Bollini and J.~J. Giambiagi.
\newblock {Dimensional Renormalization: The Number of Dimensions as a
  Regularizing Parameter}.
\newblock {\em Nuovo Cim. B}, 12:20--26, 1972.
\newblock \href {https://doi.org/10.1007/BF02895558}
  {\path{doi:10.1007/BF02895558}}.

\bibitem{Gastmans:1973uv}
R.~Gastmans and R.~Meuldermans.
\newblock {Dimensional regularization of the infrared problem}.
\newblock {\em Nucl. Phys. B}, 63:277--284, 1973.
\newblock \href {https://doi.org/10.1016/0550-3213(73)90146-6}
  {\path{doi:10.1016/0550-3213(73)90146-6}}.

\bibitem{Marciano:1975de}
W.~J. Marciano.
\newblock {Dimensional Regularization and Mass Singularities}.
\newblock {\em Phys. Rev. D}, 12:3861, 1975.
\newblock \href {https://doi.org/10.1103/PhysRevD.12.3861}
  {\path{doi:10.1103/PhysRevD.12.3861}}.

\bibitem{Altarelli:1979ub}
Guido Altarelli, R.~Keith Ellis, and G.~Martinelli.
\newblock {Large Perturbative Corrections to the Drell-Yan Process in QCD}.
\newblock {\em Nucl. Phys. B}, 157:461--497, 1979.
\newblock \href {https://doi.org/10.1016/0550-3213(79)90116-0}
  {\path{doi:10.1016/0550-3213(79)90116-0}}.

\bibitem{Kubar-Andre:1978eri}
J.~Kubar-Andre and Frank~E. Paige.
\newblock {Gluon Corrections to the Drell-Yan Model}.
\newblock {\em Phys. Rev. D}, 19:221, 1979.
\newblock \href {https://doi.org/10.1103/PhysRevD.19.221}
  {\path{doi:10.1103/PhysRevD.19.221}}.

\bibitem{Altarelli:1978id}
Guido Altarelli, R.~Keith Ellis, and G.~Martinelli.
\newblock {Leptoproduction and Drell-Yan Processes Beyond the Leading
  Approximation in Chromodynamics}.
\newblock {\em Nucl. Phys. B}, 143:521, 1978.
\newblock [Erratum: Nucl.Phys.B 146, 544 (1978)].
\newblock \href {https://doi.org/10.1016/0550-3213(78)90067-6}
  {\path{doi:10.1016/0550-3213(78)90067-6}}.

\bibitem{Harada:1979bj}
K.~Harada, T.~Kaneko, and N.~Sakai.
\newblock {Hadronic Lepton Pair Production Beyond the Leading Order in
  Perturbative QCD}.
\newblock {\em Nucl. Phys. B}, 155:169--188, 1979.
\newblock [Erratum: Nucl.Phys.B 165, 545 (1980)].
\newblock \href {https://doi.org/10.1016/0550-3213(79)90361-4}
  {\path{doi:10.1016/0550-3213(79)90361-4}}.

\bibitem{Abad:1978ke}
J.~Abad and B.~Humpert.
\newblock {Perturbative {QCD} Corrections in {Drell-Yan} Processes}.
\newblock {\em Phys. Lett. B}, 80:286--289, 1979.
\newblock \href {https://doi.org/10.1016/0370-2693(79)90218-1}
  {\path{doi:10.1016/0370-2693(79)90218-1}}.

\bibitem{Saclay-CERN-CollegedeFrance-EcolePoly-Orsay:1979yqw}
J.~Badier et~al.
\newblock {Experimental Cross-Section for Dimuon Production and the Drell-Yan
  Model}.
\newblock {\em Phys. Lett. B}, 89:145, 1979.
\newblock \href {https://doi.org/10.1016/0370-2693(79)90093-5}
  {\path{doi:10.1016/0370-2693(79)90093-5}}.

\bibitem{Altarelli:1979zh}
Guido Altarelli.
\newblock {Lepton Pair Production in Hadronic Collisions}.
\newblock In {\em {1979 EPS High-Energy Physics Conference}}, volume~2, pages
  727--741, Geneva, Switzerland, 1979. CERN.

\bibitem{Floratos:1977au}
E.~G. Floratos, D.~A. Ross, and Christopher~T. Sachrajda.
\newblock {Higher Order Effects in Asymptotically Free Gauge Theories: The
  Anomalous Dimensions of Wilson Operators}.
\newblock {\em Nucl. Phys. B}, 129:66--88, 1977.
\newblock [Erratum: Nucl.Phys.B 139, 545--546 (1978)].
\newblock \href {https://doi.org/10.1016/0550-3213(77)90020-7}
  {\path{doi:10.1016/0550-3213(77)90020-7}}.

\bibitem{Floratos:1978ny}
E.~G. Floratos, D.~A. Ross, and Christopher~T. Sachrajda.
\newblock {Higher Order Effects in Asymptotically Free Gauge Theories. 2.
  Flavor Singlet Wilson Operators and Coefficient Functions}.
\newblock {\em Nucl. Phys. B}, 152:493--520, 1979.
\newblock \href {https://doi.org/10.1016/0550-3213(79)90094-4}
  {\path{doi:10.1016/0550-3213(79)90094-4}}.

\bibitem{Bardeen:1978yd}
William~A. Bardeen, A.~J. Buras, D.~W. Duke, and T.~Muta.
\newblock {Deep Inelastic Scattering Beyond the Leading Order in Asymptotically
  Free Gauge Theories}.
\newblock {\em Phys. Rev. D}, 18:3998, 1978.
\newblock \href {https://doi.org/10.1103/PhysRevD.18.3998}
  {\path{doi:10.1103/PhysRevD.18.3998}}.

\bibitem{DeRujula:1976baf}
Alvaro De~Rujula, Howard Georgi, and H.~David Politzer.
\newblock {Demythification of Electroproduction, Local Duality and Precocious
  Scaling}.
\newblock {\em Annals Phys.}, 103:315, 1977.
\newblock \href {https://doi.org/10.1016/S0003-4916(97)90003-8}
  {\path{doi:10.1016/S0003-4916(97)90003-8}}.

\bibitem{Calvo:1977ba}
M.~Calvo.
\newblock {Higher Order Effects of Asymptotically Free Field Theories}.
\newblock {\em Phys. Rev. D}, 15:730--754, 1977.
\newblock \href {https://doi.org/10.1103/PhysRevD.15.730}
  {\path{doi:10.1103/PhysRevD.15.730}}.

\bibitem{Ellis:1978sf}
R.~Keith Ellis, Howard Georgi, Marie Machacek, H.~David Politzer, and Graham~G.
  Ross.
\newblock {Factorization and the Parton Model in QCD}.
\newblock {\em Phys. Lett. B}, 78:281--284, 1978.
\newblock \href {https://doi.org/10.1016/0370-2693(78)90023-0}
  {\path{doi:10.1016/0370-2693(78)90023-0}}.

\bibitem{Ellis:1978ty}
R.~Keith Ellis, Howard Georgi, Marie Machacek, H.~David Politzer, and Graham~G.
  Ross.
\newblock {Perturbation Theory and the Parton Model in QCD}.
\newblock {\em Nucl. Phys. B}, 152:285--329, 1979.
\newblock \href {https://doi.org/10.1016/0550-3213(79)90105-6}
  {\path{doi:10.1016/0550-3213(79)90105-6}}.

\bibitem{Amati:1978by}
D.~Amati, R.~Petronzio, and G.~Veneziano.
\newblock {Relating Hard QCD Processes Through Universality of Mass
  Singularities. 2.}
\newblock {\em Nucl. Phys. B}, 146:29--49, 1978.
\newblock \href {https://doi.org/10.1016/0550-3213(78)90430-3}
  {\path{doi:10.1016/0550-3213(78)90430-3}}.

\bibitem{Collins:1982wa}
John~C. Collins, Davison~E. Soper, and George~F. Sterman.
\newblock {Factorization for One Loop Corrections in the {Drell-Yan} Process}.
\newblock {\em Nucl. Phys. B}, 223:381--421, 1983.
\newblock \href {https://doi.org/10.1016/0550-3213(83)90062-7}
  {\path{doi:10.1016/0550-3213(83)90062-7}}.

\bibitem{Collins:1983ju}
John~C. Collins, Davison~E. Soper, and George~F. Sterman.
\newblock {All Order Factorization for {Drell-Yan} Cross-sections}.
\newblock {\em Phys. Lett. B}, 134:263, 1984.
\newblock \href {https://doi.org/10.1016/0370-2693(84)90684-1}
  {\path{doi:10.1016/0370-2693(84)90684-1}}.

\bibitem{Collins:1988ig}
John~C. Collins, Davison~E. Soper, and George~F. Sterman.
\newblock {Soft Gluons and Factorization}.
\newblock {\em Nucl. Phys. B}, 308:833--856, 1988.
\newblock \href {https://doi.org/10.1016/0550-3213(88)90130-7}
  {\path{doi:10.1016/0550-3213(88)90130-7}}.

\bibitem{Collins:1985ue}
John~C. Collins, Davison~E. Soper, and George~F. Sterman.
\newblock {Factorization for Short Distance Hadron - Hadron Scattering}.
\newblock {\em Nucl. Phys. B}, 261:104--142, 1985.
\newblock \href {https://doi.org/10.1016/0550-3213(85)90565-6}
  {\path{doi:10.1016/0550-3213(85)90565-6}}.

\bibitem{Collins:1989gx}
John~C. Collins, Davison~E. Soper, and George~F. Sterman.
\newblock {Factorization of Hard Processes in QCD}.
\newblock {\em Adv. Ser. Direct. High Energy Phys.}, 5:1--91, 1989.
\newblock \href {https://arxiv.org/abs/hep-ph/0409313}
  {\path{arXiv:hep-ph/0409313}}, \href
  {https://doi.org/10.1142/9789814503266\_0001}
  {\path{doi:10.1142/9789814503266\_0001}}.

\bibitem{Curci:1980uw}
G.~Curci, W.~Furmanski, and R.~Petronzio.
\newblock {Evolution of Parton Densities Beyond Leading Order: The Nonsinglet
  Case}.
\newblock {\em Nucl. Phys. B}, 175:27--92, 1980.
\newblock \href {https://doi.org/10.1016/0550-3213(80)90003-6}
  {\path{doi:10.1016/0550-3213(80)90003-6}}.

\bibitem{Furmanski:1980cm}
W.~Furmanski and R.~Petronzio.
\newblock {Singlet Parton Densities Beyond Leading Order}.
\newblock {\em Phys. Lett. B}, 97:437--442, 1980.
\newblock \href {https://doi.org/10.1016/0370-2693(80)90636-X}
  {\path{doi:10.1016/0370-2693(80)90636-X}}.

\bibitem{Ellis:1996nn}
R.~Keith Ellis and W.~Vogelsang.
\newblock {The Evolution of parton distributions beyond leading order: The
  Singlet case}.
\newblock 2 1996.
\newblock \href {https://arxiv.org/abs/hep-ph/9602356}
  {\path{arXiv:hep-ph/9602356}}.

\bibitem{Hamberg:1991qt}
R.~Hamberg and W.~L. van Neerven.
\newblock {The Correct renormalization of the gluon operator in a covariant
  gauge}.
\newblock {\em Nucl. Phys. B}, 379:143--171, 1992.
\newblock \href {https://doi.org/10.1016/0550-3213(92)90593-Z}
  {\path{doi:10.1016/0550-3213(92)90593-Z}}.

\bibitem{Abramowicz:1981be}
H.~Abramowicz et~al.
\newblock {Determination of the Gluon Distribution in the Nucleon from Deep
  Inelastic Neutrino Scattering}.
\newblock {\em Z. Phys. C}, 12:289, 1982.
\newblock \href {https://doi.org/10.1007/BF01557574}
  {\path{doi:10.1007/BF01557574}}.

\bibitem{Hanson:1975fe}
G.~Hanson et~al.
\newblock {Evidence for Jet Structure in Hadron Production by $e^+e^-$
  Annihilation}.
\newblock {\em Phys. Rev. Lett.}, 35:1609--1612, 1975.
\newblock \href {https://doi.org/10.1103/PhysRevLett.35.1609}
  {\path{doi:10.1103/PhysRevLett.35.1609}}.

\bibitem{Ellis:1976uc}
John~R. Ellis, Mary~K. Gaillard, and Graham~G. Ross.
\newblock {Search for Gluons in $e^+e^-$ Annihilation}.
\newblock {\em Nucl. Phys. B}, 111:253, 1976.
\newblock [Erratum: Nucl.Phys.B 130, 516 (1977)].
\newblock \href {https://doi.org/10.1016/0550-3213(77)90253-X}
  {\path{doi:10.1016/0550-3213(77)90253-X}}.

\bibitem{Sterman:1977wj}
George~F. Sterman and Steven Weinberg.
\newblock {Jets from Quantum Chromodynamics}.
\newblock {\em Phys. Rev. Lett.}, 39:1436, 1977.
\newblock \href {https://doi.org/10.1103/PhysRevLett.39.1436}
  {\path{doi:10.1103/PhysRevLett.39.1436}}.

\bibitem{Georgi:1977sf}
Howard Georgi and Marie Machacek.
\newblock {A Simple QCD Prediction of Jet Structure in $e^+e^-$ Annihilation}.
\newblock {\em Phys. Rev. Lett.}, 39:1237, 1977.
\newblock \href {https://doi.org/10.1103/PhysRevLett.39.1237}
  {\path{doi:10.1103/PhysRevLett.39.1237}}.

\bibitem{Parisi:1978eg}
G.~Parisi.
\newblock {Super Inclusive Cross-Sections}.
\newblock {\em Phys. Lett. B}, 74:65--67, 1978.
\newblock \href {https://doi.org/10.1016/0370-2693(78)90061-8}
  {\path{doi:10.1016/0370-2693(78)90061-8}}.

\bibitem{Farhi:1977sg}
Edward Farhi.
\newblock {A QCD Test for Jets}.
\newblock {\em Phys. Rev. Lett.}, 39:1587--1588, 1977.
\newblock \href {https://doi.org/10.1103/PhysRevLett.39.1587}
  {\path{doi:10.1103/PhysRevLett.39.1587}}.

\bibitem{DELPHI:1992qrr}
P.~Abreu et~al.
\newblock {Determination of alpha(s) in second order QCD from hadronic Z
  decays}.
\newblock {\em Z. Phys. C}, 54:55--74, 1992.
\newblock \href {https://doi.org/10.1007/BF01881708}
  {\path{doi:10.1007/BF01881708}}.

\bibitem{Dokshitzer:1978yd}
Yuri~L. Dokshitzer, Dmitri Diakonov, and S.~I. Troian.
\newblock {On the Transverse Momentum Distribution of Massive Lepton Pairs}.
\newblock {\em Phys. Lett. B}, 79:269--272, 1978.
\newblock \href {https://doi.org/10.1016/0370-2693(78)90240-X}
  {\path{doi:10.1016/0370-2693(78)90240-X}}.

\bibitem{Parisi:1979se}
G.~Parisi and R.~Petronzio.
\newblock {Small Transverse Momentum Distributions in Hard Processes}.
\newblock {\em Nucl. Phys. B}, 154:427--440, 1979.
\newblock \href {https://doi.org/10.1016/0550-3213(79)90040-3}
  {\path{doi:10.1016/0550-3213(79)90040-3}}.

\bibitem{Collins:1984kg}
John~C. Collins, Davison~E. Soper, and George~F. Sterman.
\newblock {Transverse Momentum Distribution in Drell-Yan Pair and W and Z Boson
  Production}.
\newblock {\em Nucl. Phys. B}, 250:199--224, 1985.
\newblock \href {https://doi.org/10.1016/0550-3213(85)90479-1}
  {\path{doi:10.1016/0550-3213(85)90479-1}}.

\bibitem{Altarelli:1984kp}
Guido Altarelli, R.~Keith Ellis, and G.~Martinelli.
\newblock {Vector Boson Production at Present and Future Colliders}.
\newblock {\em Z. Phys. C}, 27:617, 1985.
\newblock \href {https://doi.org/10.1007/BF01436519}
  {\path{doi:10.1007/BF01436519}}.

\bibitem{Altarelli:1984pt}
Guido Altarelli, R.~Keith Ellis, Mario Greco, and G.~Martinelli.
\newblock {Vector Boson Production at Colliders: A Theoretical Reappraisal}.
\newblock {\em Nucl. Phys. B}, 246:12--44, 1984.
\newblock \href {https://doi.org/10.1016/0550-3213(84)90112-3}
  {\path{doi:10.1016/0550-3213(84)90112-3}}.

\bibitem{Dasgupta:2001sh}
M.~Dasgupta and G.~P. Salam.
\newblock {Resummation of nonglobal QCD observables}.
\newblock {\em Phys. Lett. B}, 512:323--330, 2001.
\newblock \href {https://arxiv.org/abs/hep-ph/0104277}
  {\path{arXiv:hep-ph/0104277}}, \href
  {https://doi.org/10.1016/S0370-2693(01)00725-0}
  {\path{doi:10.1016/S0370-2693(01)00725-0}}.

\bibitem{Ellis:1980wv}
R.~Keith Ellis, D.~A. Ross, and A.~E. Terrano.
\newblock {The Perturbative Calculation of Jet Structure in $e^+e^-$
  Annihilation}.
\newblock {\em Nucl. Phys. B}, 178:421--456, 1981.
\newblock \href {https://doi.org/10.1016/0550-3213(81)90165-6}
  {\path{doi:10.1016/0550-3213(81)90165-6}}.

\bibitem{Kunszt:1989km}
Z.~Kunszt, P.~Nason, G.~Marchesini, and B.~R. Webber.
\newblock {QCD at LEP, available at \url{"10.5170/CERN-1989-008-V-1.373"}}.
\newblock In {\em {LEP Physics Workshop}}, 8 1989.

\bibitem{Frixione:1995ms}
S.~Frixione, Z.~Kunszt, and A.~Signer.
\newblock {Three jet cross-sections to next-to-leading order}.
\newblock {\em Nucl. Phys. B}, 467:399--442, 1996.
\newblock \href {https://arxiv.org/abs/hep-ph/9512328}
  {\path{arXiv:hep-ph/9512328}}, \href
  {https://doi.org/10.1016/0550-3213(96)00110-1}
  {\path{doi:10.1016/0550-3213(96)00110-1}}.

\bibitem{Catani:1996vz}
S.~Catani and M.~H. Seymour.
\newblock {A General algorithm for calculating jet cross-sections in NLO QCD}.
\newblock {\em Nucl. Phys. B}, 485:291--419, 1997.
\newblock [Erratum: Nucl.Phys.B 510, 503--504 (1998)].
\newblock \href {https://arxiv.org/abs/hep-ph/9605323}
  {\path{arXiv:hep-ph/9605323}}, \href
  {https://doi.org/10.1016/S0550-3213(96)00589-5}
  {\path{doi:10.1016/S0550-3213(96)00589-5}}.

\bibitem{Fabricius:1981sx}
K.~Fabricius, I.~Schmitt, G.~Kramer, and G.~Schierholz.
\newblock {Higher Order Perturbative QCD Calculation of Jet Cross-Sections in
  e+ e- Annihilation}.
\newblock {\em Z. Phys. C}, 11:315, 1981.
\newblock \href {https://doi.org/10.1007/BF01578281}
  {\path{doi:10.1007/BF01578281}}.

\bibitem{Giele:1991vf}
W.~T. Giele and E.~W.~Nigel Glover.
\newblock {Higher order corrections to jet cross-sections in e+ e-
  annihilation}.
\newblock {\em Phys. Rev. D}, 46:1980--2010, 1992.
\newblock \href {https://doi.org/10.1103/PhysRevD.46.1980}
  {\path{doi:10.1103/PhysRevD.46.1980}}.

\bibitem{Veltman}
M.~Veltman.
\newblock Schoonschip history, available at
  \url{"https://vsys.physics.lsa.umich.edu/schip-docs/CERN-Schoonschip-1967.pdf"},
  1967.

\bibitem{Passarino:1978jh}
G.~Passarino and M.~J.~G. Veltman.
\newblock {One Loop Corrections for $e^+e^-$ Annihilation Into mu+ mu- in the
  Weinberg Model}.
\newblock {\em Nucl. Phys. B}, 160:151--207, 1979.
\newblock \href {https://doi.org/10.1016/0550-3213(79)90234-7}
  {\path{doi:10.1016/0550-3213(79)90234-7}}.

\bibitem{tHooft:1978jhc}
Gerard 't~Hooft and M.~J.~G. Veltman.
\newblock {Scalar One Loop Integrals}.
\newblock {\em Nucl. Phys. B}, 153:365--401, 1979.
\newblock \href {https://doi.org/10.1016/0550-3213(79)90605-9}
  {\path{doi:10.1016/0550-3213(79)90605-9}}.

\bibitem{vanOldenborgh:1990yc}
G.~J. van Oldenborgh.
\newblock {FF: A Package to evaluate one loop Feynman diagrams}.
\newblock {\em Comput. Phys. Commun.}, 66:1--15, 1991.
\newblock \href {https://doi.org/10.1016/0010-4655(91)90002-3}
  {\path{doi:10.1016/0010-4655(91)90002-3}}.

\bibitem{Hahn:1998yk}
T.~Hahn and M.~Perez-Victoria.
\newblock {Automatized one loop calculations in four-dimensions and
  D-dimensions}.
\newblock {\em Comput. Phys. Commun.}, 118:153--165, 1999.
\newblock \href {https://arxiv.org/abs/hep-ph/9807565}
  {\path{arXiv:hep-ph/9807565}}, \href
  {https://doi.org/10.1016/S0010-4655(98)00173-8}
  {\path{doi:10.1016/S0010-4655(98)00173-8}}.

\bibitem{Ellis:2007qk}
R.~Keith Ellis and Giulia Zanderighi.
\newblock {Scalar one-loop integrals for QCD}.
\newblock {\em JHEP}, 02:002, 2008.
\newblock \href {https://arxiv.org/abs/0712.1851} {\path{arXiv:0712.1851}},
  \href {https://doi.org/10.1088/1126-6708/2008/02/002}
  {\path{doi:10.1088/1126-6708/2008/02/002}}.

\bibitem{Kublbeck:1990xc}
J.~Kublbeck, M.~Bohm, and Ansgar Denner.
\newblock {Feyn Arts: Computer Algebraic Generation of Feynman Graphs and
  Amplitudes}.
\newblock {\em Comput. Phys. Commun.}, 60:165--180, 1990.
\newblock \href {https://doi.org/10.1016/0010-4655(90)90001-H}
  {\path{doi:10.1016/0010-4655(90)90001-H}}.

\bibitem{Mertig:1990an}
R.~Mertig, M.~Bohm, and Ansgar Denner.
\newblock {FEYN CALC: Computer algebraic calculation of Feynman amplitudes}.
\newblock {\em Comput. Phys. Commun.}, 64:345--359, 1991.
\newblock \href {https://doi.org/10.1016/0010-4655(91)90130-D}
  {\path{doi:10.1016/0010-4655(91)90130-D}}.

\bibitem{Vermaseren:1992vn}
J.~A.~M. Vermaseren.
\newblock {The Symbolic manipulation program FORM, available at
  \url{"https://lib-extopc.kek.jp/preprints/PDF/1993/9303/9303412.pdf"}}.
\newblock 3 1992.

\bibitem{Vermaseren:2000nd}
J.~A.~M. Vermaseren.
\newblock {New features of FORM}.
\newblock 10 2000.
\newblock \href {https://arxiv.org/abs/math-ph/0010025}
  {\path{arXiv:math-ph/0010025}}.

\bibitem{Caswell:1974gg}
William~E. Caswell.
\newblock {Asymptotic Behavior of Nonabelian Gauge Theories to Two Loop Order}.
\newblock {\em Phys. Rev. Lett.}, 33:244, 1974.
\newblock \href {https://doi.org/10.1103/PhysRevLett.33.244}
  {\path{doi:10.1103/PhysRevLett.33.244}}.

\bibitem{Jones:1974mm}
D.~R.~T. Jones.
\newblock {Two Loop Diagrams in Yang-Mills Theory}.
\newblock {\em Nucl. Phys. B}, 75:531, 1974.
\newblock \href {https://doi.org/10.1016/0550-3213(74)90093-5}
  {\path{doi:10.1016/0550-3213(74)90093-5}}.

\bibitem{Tarasov:1980au}
O.~V. Tarasov, A.~A. Vladimirov, and A.~Yu. Zharkov.
\newblock {The Gell-Mann-Low Function of QCD in the Three Loop Approximation}.
\newblock {\em Phys. Lett. B}, 93:429--432, 1980.
\newblock \href {https://doi.org/10.1016/0370-2693(80)90358-5}
  {\path{doi:10.1016/0370-2693(80)90358-5}}.

\bibitem{Larin:1993tp}
S.~A. Larin and J.~A.~M. Vermaseren.
\newblock {The Three loop QCD Beta function and anomalous dimensions}.
\newblock {\em Phys. Lett. B}, 303:334--336, 1993.
\newblock \href {https://arxiv.org/abs/hep-ph/9302208}
  {\path{arXiv:hep-ph/9302208}}, \href
  {https://doi.org/10.1016/0370-2693(93)91441-O}
  {\path{doi:10.1016/0370-2693(93)91441-O}}.

\bibitem{vanRitbergen:1997va}
T.~van Ritbergen, J.~A.~M. Vermaseren, and S.~A. Larin.
\newblock {The Four loop beta function in quantum chromodynamics}.
\newblock {\em Phys. Lett. B}, 400:379--384, 1997.
\newblock \href {https://arxiv.org/abs/hep-ph/9701390}
  {\path{arXiv:hep-ph/9701390}}, \href
  {https://doi.org/10.1016/S0370-2693(97)00370-5}
  {\path{doi:10.1016/S0370-2693(97)00370-5}}.

\bibitem{Baikov:2016tgj}
P.~A. Baikov, K.~G. Chetyrkin, and J.~H. K{\"u}hn.
\newblock {Five-Loop Running of the QCD Coupling Constant}.
\newblock {\em Phys. Rev. Lett.}, 118(8):082002, 2017.
\newblock \href {https://arxiv.org/abs/1606.08659} {\path{arXiv:1606.08659}},
  \href {https://doi.org/10.1103/PhysRevLett.118.082002}
  {\path{doi:10.1103/PhysRevLett.118.082002}}.

\bibitem{Herzog:2017ohr}
F.~Herzog, B.~Ruijl, T.~Ueda, J.~A.~M. Vermaseren, and A.~Vogt.
\newblock {The five-loop beta function of Yang-Mills theory with fermions}.
\newblock {\em JHEP}, 02:090, 2017.
\newblock \href {https://arxiv.org/abs/1701.01404} {\path{arXiv:1701.01404}},
  \href {https://doi.org/10.1007/JHEP02(2017)090}
  {\path{doi:10.1007/JHEP02(2017)090}}.

\bibitem{Herzog:2025ngd}
Franz Herzog, Ben Ruijl, Takahiro Ueda, Jos Vermaseren, and Andreas Vogt.
\newblock {Five-loop beta function for gauge theories: computations, results
  and consequences}.
\newblock 10 2025.
\newblock \href {https://arxiv.org/abs/2510.21624} {\path{arXiv:2510.21624}}.

\bibitem{Schwartz:2014sze}
Matthew~D. Schwartz.
\newblock {\em {Quantum Field Theory and the Standard Model}}.
\newblock Cambridge University Press, 3 2014.
\newblock \href {https://doi.org/10.1017/9781139540940}
  {\path{doi:10.1017/9781139540940}}.

\bibitem{Berends:1981rb}
Frits~A. Berends, R.~Kleiss, P.~De~Causmaecker, R.~Gastmans, and Tai~Tsun Wu.
\newblock {Single Bremsstrahlung Processes in Gauge Theories}.
\newblock {\em Phys. Lett. B}, 103:124--128, 1981.
\newblock \href {https://doi.org/10.1016/0370-2693(81)90685-7}
  {\path{doi:10.1016/0370-2693(81)90685-7}}.

\bibitem{Berends:1981uq}
Frits~A. Berends, R.~Kleiss, P.~De~Causmaecker, R.~Gastmans, W.~Troost, and
  Tai~Tsun Wu.
\newblock {Multiple Bremsstrahlung in Gauge Theories at High-Energies. 2.
  Single Bremsstrahlung}.
\newblock {\em Nucl. Phys. B}, 206:61--89, 1982.
\newblock \href {https://doi.org/10.1016/0550-3213(82)90489-8}
  {\path{doi:10.1016/0550-3213(82)90489-8}}.

\bibitem{DeCausmaecker:1981jtq}
P.~De~Causmaecker, R.~Gastmans, W.~Troost, and Tai~Tsun Wu.
\newblock {Multiple Bremsstrahlung in Gauge Theories at High-Energies. 1.
  General Formalism for Quantum Electrodynamics}.
\newblock {\em Nucl. Phys. B}, 206:53--60, 1982.
\newblock \href {https://doi.org/10.1016/0550-3213(82)90488-6}
  {\path{doi:10.1016/0550-3213(82)90488-6}}.

\bibitem{Xu:1984qe}
Zhan Xu, Da-Hua Zhang, and Lee Chang.
\newblock {Helicity amplitudes for multiple bremsstrahlung in massless
  nonabelian gauge theory. 1. New definition of polarization vector and
  formulation of amplitudes in Grassmann algebra, available at
  \url{https://inspirehep.net/files/a7766622994a11f71eb1d71d0a93f69d}}.
\newblock 12 1984.

\bibitem{Xu:1985cs}
Zhan Xu, Da-Hua Zhang, and Lee Chang.
\newblock {Helicity amplitudes for multiple bremsstrahlung in massless
  nonabelian gauge theory. 2. Decomposition of amplitudes into gauge invariant
  subsets, available at
  \url{https://inspirehep.net/files/a7766622994a11f71eb1d71d0a93f69d}}.
\newblock 3 1985.

\bibitem{Zhang:1985nk}
Da-Hua Zhang, Zhan Xu, and Lee Chang.
\newblock {Helicity amplitudes for multiple bremsstrahlung in massless
  nonabelian gauge theory. 3. Amplitudes of multiple bremsstrahlung processes
  involving gluon selfcoupling vertices, available at
  \url{https://inspirehep.net/files/a7766622994a11f71eb1d71d0a93f69d}}.
\newblock 1 1985.

\bibitem{Xu:1986xb}
Zhan Xu, Da-Hua Zhang, and Lee Chang.
\newblock {Helicity Amplitudes for Multiple Bremsstrahlung in Massless
  Nonabelian Gauge Theories}.
\newblock {\em Nucl. Phys. B}, 291:392--428, 1987.
\newblock \href {https://doi.org/10.1016/0550-3213(87)90479-2}
  {\path{doi:10.1016/0550-3213(87)90479-2}}.

\bibitem{Parke:1986gb}
Stephen~J. Parke and T.~R. Taylor.
\newblock {An Amplitude for $n$ Gluon Scattering}.
\newblock {\em Phys. Rev. Lett.}, 56:2459, 1986.
\newblock \href {https://doi.org/10.1103/PhysRevLett.56.2459}
  {\path{doi:10.1103/PhysRevLett.56.2459}}.

\bibitem{Berends:1987me}
Frits~A. Berends and W.~T. Giele.
\newblock {Recursive Calculations for Processes with n Gluons}.
\newblock {\em Nucl. Phys. B}, 306:759--808, 1988.
\newblock \href {https://doi.org/10.1016/0550-3213(88)90442-7}
  {\path{doi:10.1016/0550-3213(88)90442-7}}.

\bibitem{Elvang:2013cua}
Henriette Elvang and Yu-tin Huang.
\newblock {Scattering Amplitudes}.
\newblock 8 2013.
\newblock \href {https://arxiv.org/abs/1308.1697} {\path{arXiv:1308.1697}}.

\bibitem{Vermaseren:1980qz}
J.~A.~M. Vermaseren, K.~J.~F. Gaemers, and S.~J. Oldham.
\newblock {Perturbative QCD Calculation of Jet Cross-Sections in $e^+e^-$
  Annihilation}.
\newblock {\em Nucl. Phys. B}, 187:301--320, 1981.
\newblock \href {https://doi.org/10.1016/0550-3213(81)90276-5}
  {\path{doi:10.1016/0550-3213(81)90276-5}}.

\bibitem{Kunszt:1980vt}
Zoltan Kunszt.
\newblock {Comment on the O($\alpha^2_s$) Corrections to Jet Production in
  $e^+e^-$ Annihilation}.
\newblock {\em Phys. Lett. B}, 99:429--432, 1981.
\newblock \href {https://doi.org/10.1016/0370-2693(81)90563-3}
  {\path{doi:10.1016/0370-2693(81)90563-3}}.

\bibitem{Nachtmann:1982xr}
O.~Nachtmann and A.~Reiter.
\newblock {A Test for the Gluon Selfcoupling in the Reactions $e^+e^-$
  ---{\ensuremath{>}} Four Jets and Z0 ---{\ensuremath{>}} Four Jets}.
\newblock {\em Z. Phys. C}, 16:45, 1982.
\newblock \href {https://doi.org/10.1007/BF01573746}
  {\path{doi:10.1007/BF01573746}}.

\bibitem{Bengtsson:1988qg}
M.~Bengtsson and P.~M. Zerwas.
\newblock {Four Jet Events in $e^+e^-$ Annihilation: Testing the Three Gluon
  Vertex}.
\newblock {\em Phys. Lett. B}, 208:306--308, 1988.
\newblock \href {https://doi.org/10.1016/0370-2693(88)90435-2}
  {\path{doi:10.1016/0370-2693(88)90435-2}}.

\bibitem{Bern:1997sc}
Zvi Bern, Lance~J. Dixon, and David~A. Kosower.
\newblock {One loop amplitudes for e+ e- to four partons}.
\newblock {\em Nucl. Phys. B}, 513:3--86, 1998.
\newblock \href {https://arxiv.org/abs/hep-ph/9708239}
  {\path{arXiv:hep-ph/9708239}}, \href
  {https://doi.org/10.1016/S0550-3213(97)00703-7}
  {\path{doi:10.1016/S0550-3213(97)00703-7}}.

\bibitem{Dixon:1997th}
Lance~J. Dixon and Adrian Signer.
\newblock {Complete O ($\alpha_s^3$) results for $e^+ e^- \to (\gamma, Z) \to$~
  four jets}.
\newblock {\em Phys. Rev. D}, 56:4031--4038, 1997.
\newblock \href {https://arxiv.org/abs/hep-ph/9706285}
  {\path{arXiv:hep-ph/9706285}}, \href
  {https://doi.org/10.1103/PhysRevD.56.4031}
  {\path{doi:10.1103/PhysRevD.56.4031}}.

\bibitem{Catani:1991hj}
S.~Catani, Yuri~L. Dokshitzer, M.~Olsson, G.~Turnock, and B.~R. Webber.
\newblock {New clustering algorithm for multi - jet cross-sections in e+ e-
  annihilation}.
\newblock {\em Phys. Lett. B}, 269:432--438, 1991.
\newblock \href {https://doi.org/10.1016/0370-2693(91)90196-W}
  {\path{doi:10.1016/0370-2693(91)90196-W}}.

\bibitem{Buras:1977yj}
A.~J. Buras and K.~J.~F. Gaemers.
\newblock {Simple Parametrizations of Parton Distributions with q**2 Dependence
  Given by Asymptotic Freedom}.
\newblock {\em Nucl. Phys. B}, 132:249--267, 1978.
\newblock \href {https://doi.org/10.1016/0550-3213(78)90119-0}
  {\path{doi:10.1016/0550-3213(78)90119-0}}.

\bibitem{Duke:1983gd}
D.~W. Duke and J.~F. Owens.
\newblock {Q**2 Dependent Parametrizations of Parton Distribution Functions}.
\newblock {\em Phys. Rev. D}, 30:49--54, 1984.
\newblock \href {https://doi.org/10.1103/PhysRevD.30.49}
  {\path{doi:10.1103/PhysRevD.30.49}}.

\bibitem{Martin:1987vw}
Alan~D. Martin, R.~G. Roberts, and W.~James Stirling.
\newblock {Structure Function Analysis and psi, Jet, W, Z Production: Pinning
  Down the Gluon}.
\newblock {\em Phys. Rev. D}, 37:1161, 1988.
\newblock \href {https://doi.org/10.1103/PhysRevD.37.1161}
  {\path{doi:10.1103/PhysRevD.37.1161}}.

\bibitem{Botts:1993shg}
James Botts, Jorge~G. Morfin, Joseph~F. Owens, Jian-wei Qiu, Wu-Ki Tung, and
  Harry Weerts.
\newblock {CTEQ parton distributions and flavor dependence of sea quarks}.
\newblock {\em Phys. Lett. B}, 304:159--166, 1993.
\newblock \href {https://arxiv.org/abs/hep-ph/9303255}
  {\path{arXiv:hep-ph/9303255}}, \href
  {https://doi.org/10.1016/0370-2693(93)91416-K}
  {\path{doi:10.1016/0370-2693(93)91416-K}}.

\bibitem{Martin:1988nk}
Alan~D. Martin, R.~G. Roberts, and W.~James Stirling.
\newblock {Implications of New Deep Inelastic Scattering Data for Parton
  Distributions}.
\newblock {\em Phys. Lett. B}, 206:327--332, 1988.
\newblock \href {https://doi.org/10.1016/0370-2693(88)91515-8}
  {\path{doi:10.1016/0370-2693(88)91515-8}}.

\bibitem{Gluck:1991ng}
M.~Gluck, E.~Reya, and A.~Vogt.
\newblock {Parton distributions for high-energy collisions}.
\newblock {\em Z. Phys. C}, 53:127--134, 1992.
\newblock \href {https://doi.org/10.1007/BF01483880}
  {\path{doi:10.1007/BF01483880}}.

\bibitem{Martin:1996as}
Alan~D. Martin, R.~G. Roberts, and W.~James Stirling.
\newblock {Parton distributions: A Study of the new HERA data, alpha-s, the
  gluon and p anti-p jet production}.
\newblock {\em Phys. Lett. B}, 387:419--426, 1996.
\newblock \href {https://arxiv.org/abs/hep-ph/9606345}
  {\path{arXiv:hep-ph/9606345}}, \href
  {https://doi.org/10.1016/0370-2693(96)01031-3}
  {\path{doi:10.1016/0370-2693(96)01031-3}}.

\bibitem{Owens:1992hd}
Joseph~F. Owens and Wu-Ki Tung.
\newblock {Parton distribution functions of hadrons}.
\newblock {\em Ann. Rev. Nucl. Part. Sci.}, 42:291--332, 1992.
\newblock \href {https://doi.org/10.1146/annurev.ns.42.120192.001451}
  {\path{doi:10.1146/annurev.ns.42.120192.001451}}.

\bibitem{Ellis:1985er}
R.~Keith Ellis and J.~C. Sexton.
\newblock {QCD Radiative Corrections to Parton Parton Scattering}.
\newblock {\em Nucl. Phys. B}, 269:445--484, 1986.
\newblock \href {https://doi.org/10.1016/0550-3213(86)90232-4}
  {\path{doi:10.1016/0550-3213(86)90232-4}}.

\bibitem{Ellis:1990ek}
Stephen~D. Ellis, Zoltan Kunszt, and Davison~E. Soper.
\newblock {The One Jet Inclusive Cross-section at Order $\alpha_{s}^{3}$ :
  Quarks and Gluons}.
\newblock {\em Phys. Rev. Lett.}, 64:2121, 1990.
\newblock \href {https://doi.org/10.1103/PhysRevLett.64.2121}
  {\path{doi:10.1103/PhysRevLett.64.2121}}.

\bibitem{CDF:1996yow}
F.~Abe et~al.
\newblock {Inclusive jet cross section in ${\bar p p}$ collisions at
  $\sqrt{s}=1.8$ TeV}.
\newblock {\em Phys. Rev. Lett.}, 77:438--443, 1996.
\newblock \href {https://arxiv.org/abs/hep-ex/9601008}
  {\path{arXiv:hep-ex/9601008}}, \href
  {https://doi.org/10.1103/PhysRevLett.77.438}
  {\path{doi:10.1103/PhysRevLett.77.438}}.

\bibitem{Bern:1990cu}
Zvi Bern and David~A. Kosower.
\newblock {Efficient calculation of one loop QCD amplitudes}.
\newblock {\em Phys. Rev. Lett.}, 66:1669--1672, 1991.
\newblock \href {https://doi.org/10.1103/PhysRevLett.66.1669}
  {\path{doi:10.1103/PhysRevLett.66.1669}}.

\bibitem{Kunszt:1993sd}
Zoltan Kunszt, Adrian Signer, and Zoltan Trocsanyi.
\newblock {One loop helicity amplitudes for all 2 ---{\ensuremath{>}} 2
  processes in QCD and N=1 supersymmetric Yang-Mills theory}.
\newblock {\em Nucl. Phys. B}, 411:397--442, 1994.
\newblock \href {https://arxiv.org/abs/hep-ph/9305239}
  {\path{arXiv:hep-ph/9305239}}, \href
  {https://doi.org/10.1016/0550-3213(94)90456-1}
  {\path{doi:10.1016/0550-3213(94)90456-1}}.

\bibitem{Bern:1993mq}
Zvi Bern, Lance~J. Dixon, and David~A. Kosower.
\newblock {One loop corrections to five gluon amplitudes}.
\newblock {\em Phys. Rev. Lett.}, 70:2677--2680, 1993.
\newblock \href {https://arxiv.org/abs/hep-ph/9302280}
  {\path{arXiv:hep-ph/9302280}}, \href
  {https://doi.org/10.1103/PhysRevLett.70.2677}
  {\path{doi:10.1103/PhysRevLett.70.2677}}.

\bibitem{Kunszt:1994nq}
Zoltan Kunszt, Adrian Signer, and Zoltan Trocsanyi.
\newblock {One loop radiative corrections to the helicity amplitudes of QCD
  processes involving four quarks and one gluon}.
\newblock {\em Phys. Lett. B}, 336:529--536, 1994.
\newblock \href {https://arxiv.org/abs/hep-ph/9405386}
  {\path{arXiv:hep-ph/9405386}}, \href
  {https://doi.org/10.1016/0370-2693(94)90568-1}
  {\path{doi:10.1016/0370-2693(94)90568-1}}.

\bibitem{Bern:1994fz}
Zvi Bern, Lance~J. Dixon, and David~A. Kosower.
\newblock {One loop corrections to two quark three gluon amplitudes}.
\newblock {\em Nucl. Phys. B}, 437:259--304, 1995.
\newblock \href {https://arxiv.org/abs/hep-ph/9409393}
  {\path{arXiv:hep-ph/9409393}}, \href
  {https://doi.org/10.1016/0550-3213(94)00542-M}
  {\path{doi:10.1016/0550-3213(94)00542-M}}.

\bibitem{Nagy:2003tz}
Zoltan Nagy.
\newblock {Next-to-leading order calculation of three jet observables in hadron
  hadron collision}.
\newblock {\em Phys. Rev. D}, 68:094002, 2003.
\newblock \href {https://arxiv.org/abs/hep-ph/0307268}
  {\path{arXiv:hep-ph/0307268}}, \href
  {https://doi.org/10.1103/PhysRevD.68.094002}
  {\path{doi:10.1103/PhysRevD.68.094002}}.

\bibitem{Collins:1985gm}
John~C. Collins, Davison~E. Soper, and George~F. Sterman.
\newblock {Heavy Particle Production in High-Energy Hadron Collisions}.
\newblock {\em Nucl. Phys. B}, 263:37, 1986.
\newblock \href {https://doi.org/10.1016/0550-3213(86)90026-X}
  {\path{doi:10.1016/0550-3213(86)90026-X}}.

\bibitem{Nason:1987xz}
P.~Nason, S.~Dawson, and R.~Keith Ellis.
\newblock {The Total Cross-Section for the Production of Heavy Quarks in
  Hadronic Collisions}.
\newblock {\em Nucl. Phys. B}, 303:607--633, 1988.
\newblock \href {https://doi.org/10.1016/0550-3213(88)90422-1}
  {\path{doi:10.1016/0550-3213(88)90422-1}}.

\bibitem{Nason:1989zy}
P.~Nason, S.~Dawson, and R.~Keith Ellis.
\newblock {The One Particle Inclusive Differential Cross-Section for Heavy
  Quark Production in Hadronic Collisions}.
\newblock {\em Nucl. Phys. B}, 327:49--92, 1989.
\newblock [Erratum: Nucl.Phys.B 335, 260--260 (1990)].
\newblock \href {https://doi.org/10.1016/0550-3213(89)90286-1}
  {\path{doi:10.1016/0550-3213(89)90286-1}}.

\bibitem{Beenakker:1988bq}
W.~Beenakker, H.~Kuijf, W.~L. van Neerven, and J.~Smith.
\newblock {QCD Corrections to Heavy Quark Production in p anti-p Collisions}.
\newblock {\em Phys. Rev. D}, 40:54--82, 1989.
\newblock \href {https://doi.org/10.1103/PhysRevD.40.54}
  {\path{doi:10.1103/PhysRevD.40.54}}.

\bibitem{Beenakker:1990maa}
W.~Beenakker, W.~L. van Neerven, R.~Meng, G.~A. Schuler, and J.~Smith.
\newblock {QCD corrections to heavy quark production in hadron hadron
  collisions}.
\newblock {\em Nucl. Phys. B}, 351:507--560, 1991.
\newblock \href {https://doi.org/10.1016/S0550-3213(05)80032-X}
  {\path{doi:10.1016/S0550-3213(05)80032-X}}.

\bibitem{Berends:1990ax}
Frits~A. Berends, H.~Kuijf, B.~Tausk, and W.~T. Giele.
\newblock {On the Production of a W and Jets at Hadron Colliders}.
\newblock {\em Nucl. Phys. B}, 357:32--64, 1991.
\newblock \href {https://doi.org/10.1016/0550-3213(91)90458-A}
  {\path{doi:10.1016/0550-3213(91)90458-A}}.

\bibitem{Campbell:2002tg}
John~M. Campbell and R.~Keith Ellis.
\newblock {Next-to-Leading Order Corrections to $W^+$ 2 jet and $Z^+$ 2 Jet
  Production at Hadron Colliders}.
\newblock {\em Phys. Rev. D}, 65:113007, 2002.
\newblock \href {https://arxiv.org/abs/hep-ph/0202176}
  {\path{arXiv:hep-ph/0202176}}, \href
  {https://doi.org/10.1103/PhysRevD.65.113007}
  {\path{doi:10.1103/PhysRevD.65.113007}}.

\bibitem{Ellis:1998fv}
R.~Keith Ellis and Sinisa Veseli.
\newblock {Strong radiative corrections to W b anti-b production in p anti-p
  collisions}.
\newblock {\em Phys. Rev. D}, 60:011501, 1999.
\newblock \href {https://arxiv.org/abs/hep-ph/9810489}
  {\path{arXiv:hep-ph/9810489}}, \href
  {https://doi.org/10.1103/PhysRevD.60.011501}
  {\path{doi:10.1103/PhysRevD.60.011501}}.

\bibitem{Dixon:1998py}
Lance~J. Dixon, Z.~Kunszt, and A.~Signer.
\newblock {Helicity amplitudes for O(alpha-s) production of $W^{+} W^{-}$,
  $W^\pm Z$, $Z Z$, $W^\pm \gamma$, or $Z \gamma$ pairs at hadron colliders}.
\newblock {\em Nucl. Phys. B}, 531:3--23, 1998.
\newblock \href {https://arxiv.org/abs/hep-ph/9803250}
  {\path{arXiv:hep-ph/9803250}}, \href
  {https://doi.org/10.1016/S0550-3213(98)00421-0}
  {\path{doi:10.1016/S0550-3213(98)00421-0}}.

\bibitem{Dixon:1999di}
Lance~J. Dixon, Z.~Kunszt, and A.~Signer.
\newblock {Vector boson pair production in hadronic collisions at order
  $\alpha_s$ : Lepton correlations and anomalous couplings}.
\newblock {\em Phys. Rev. D}, 60:114037, 1999.
\newblock \href {https://arxiv.org/abs/hep-ph/9907305}
  {\path{arXiv:hep-ph/9907305}}, \href
  {https://doi.org/10.1103/PhysRevD.60.114037}
  {\path{doi:10.1103/PhysRevD.60.114037}}.

\bibitem{Campbell:1999ah}
John~M. Campbell and R.~Keith Ellis.
\newblock {An Update on vector boson pair production at hadron colliders}.
\newblock {\em Phys. Rev. D}, 60:113006, 1999.
\newblock \href {https://arxiv.org/abs/hep-ph/9905386}
  {\path{arXiv:hep-ph/9905386}}, \href
  {https://doi.org/10.1103/PhysRevD.60.113006}
  {\path{doi:10.1103/PhysRevD.60.113006}}.

\bibitem{Matsuura:1988sm}
T.~Matsuura, S.~C. van~der Marck, and W.~L. van Neerven.
\newblock {The Calculation of the Second Order Soft and Virtual Contributions
  to the Drell-Yan Cross-Section}.
\newblock {\em Nucl. Phys. B}, 319:570--622, 1989.
\newblock \href {https://doi.org/10.1016/0550-3213(89)90620-2}
  {\path{doi:10.1016/0550-3213(89)90620-2}}.

\bibitem{vanNeerven:1991gh}
W.~L. van Neerven and E.~B. Zijlstra.
\newblock {The $O(\alpha_s^2)$ corrected Drell-Yan $K$ factor in the DIS and MS
  scheme}.
\newblock {\em Nucl. Phys. B}, 382:11--62, 1992.
\newblock [Erratum: Nucl.Phys.B 680, 513--514 (2004)].
\newblock \href {https://doi.org/10.1016/0550-3213(92)90078-P}
  {\path{doi:10.1016/0550-3213(92)90078-P}}.

\bibitem{vanNeerven:1999ca}
W.~L. van Neerven and A.~Vogt.
\newblock {NNLO evolution of deep inelastic structure functions: The Nonsinglet
  case}.
\newblock {\em Nucl. Phys. B}, 568:263--286, 2000.
\newblock \href {https://arxiv.org/abs/hep-ph/9907472}
  {\path{arXiv:hep-ph/9907472}}, \href
  {https://doi.org/10.1016/S0550-3213(99)00668-9}
  {\path{doi:10.1016/S0550-3213(99)00668-9}}.

\bibitem{Jarlskog:1990bk}
G.~Jarlskog and D.~Rein, editors.
\newblock {\em {ECFA Large Hadron Collider Workshop, Aachen, Germany, 4-9 Oct
  1990: Proceedings.2.}}, CERN Yellow Reports: Conference Proceedings, 12 1990.
\newblock \href {https://doi.org/10.5170/CERN-1990-010-V-2}
  {\path{doi:10.5170/CERN-1990-010-V-2}}.

\bibitem{Catani:1998bh}
Stefano Catani.
\newblock {The Singular behavior of QCD amplitudes at two loop order}.
\newblock {\em Phys. Lett. B}, 427:161--171, 1998.
\newblock \href {https://arxiv.org/abs/hep-ph/9802439}
  {\path{arXiv:hep-ph/9802439}}, \href
  {https://doi.org/10.1016/S0370-2693(98)00332-3}
  {\path{doi:10.1016/S0370-2693(98)00332-3}}.

\bibitem{Bern:2000dn}
Z.~Bern, Lance~J. Dixon, and D.~A. Kosower.
\newblock {A Two loop four gluon helicity amplitude in QCD}.
\newblock {\em JHEP}, 01:027, 2000.
\newblock \href {https://arxiv.org/abs/hep-ph/0001001}
  {\path{arXiv:hep-ph/0001001}}, \href
  {https://doi.org/10.1088/1126-6708/2000/01/027}
  {\path{doi:10.1088/1126-6708/2000/01/027}}.

\bibitem{Baikov:2012zn}
P.~A. Baikov, K.~G. Chetyrkin, J.~H. Kuhn, and J.~Rittinger.
\newblock {Adler Function, Sum Rules and Crewther Relation of Order
  $\mathcal{O}(\alpha^4_s)$: the Singlet Case}.
\newblock {\em Phys. Lett. B}, 714:62--65, 2012.
\newblock \href {https://arxiv.org/abs/1206.1288} {\path{arXiv:1206.1288}},
  \href {https://doi.org/10.1016/j.physletb.2012.06.052}
  {\path{doi:10.1016/j.physletb.2012.06.052}}.

\bibitem{Surguladze:1990tg}
Levan~R. Surguladze and Mark~A. Samuel.
\newblock {Total hadronic cross-section in $e^+e^-$ annihilation at the four
  loop level of perturbative QCD}.
\newblock {\em Phys. Rev. Lett.}, 66:560--563, 1991.
\newblock [Erratum: Phys.Rev.Lett. 66, 2416 (1991)].
\newblock \href {https://doi.org/10.1103/PhysRevLett.66.560}
  {\path{doi:10.1103/PhysRevLett.66.560}}.

\bibitem{SurguladzeErratum}
Levan~R. Surguladze and Mark~A. Samuel.
\newblock {Erratum to: "Total hadronic cross-section in $e^+e^-$ annihilation
  at the four loop level of perturbative QCD"}.
\newblock {\em Phys. Rev. Lett.}, 66:2416, 1991.
\newblock \href {https://doi.org/10.1103/PhysRevLett.66.2416}
  {\path{doi:10.1103/PhysRevLett.66.2416}}.

\bibitem{Gorishnii:1990vf}
S.~G. Gorishnii, A.~L. Kataev, and S.~A. Larin.
\newblock {The $O(\alpha^{3}_{s})$-corrections to
  $\sigma_{tot}(e^{+}e^{-}\rightarrow hadrons)$ and $\Gamma(\tau^{-}
  \rightarrow \nu_{\tau} + hadrons)$ in QCD}.
\newblock {\em Phys. Lett. B}, 259:144--150, 1991.
\newblock \href {https://doi.org/10.1016/0370-2693(91)90149-K}
  {\path{doi:10.1016/0370-2693(91)90149-K}}.

\end{thebibliography}
